\documentclass[journal=jacsat,manuscript=article]{achemso}
\setkeys{acs}{doi=true}

\usepackage[version=3]{mhchem} % Formula subscripts using \ce{}
\usepackage{hyperref}
\usepackage{graphicx} % da inserire nel preambolo

\usepackage{xcolor}  % To remove before submission
\usepackage{comment}  % To remove before submission
\usepackage{setspace}
\usepackage{caption}
\author{Lucrezia L. Baldo}
\author{Chiara Lionello}
\author{Edward D. Donkor}
\author{Matteo Becchi}
\author{Giovanni M. Pavan}
\email{giovanni.pavan@polito.it}
\affiliation[1]{Department of Applied Science and Technology, Politecnico di Torino, Corso Duca degli Abruzzi, 24, Torino 10129, Italy}

\title[Microstructural Plots]{Learning the information contained in the microstructure of data distributions}

\begin{document}
\begin{spacing}{1}
%\begin{tocentry}

%\end{tocentry}

\begin{abstract}
Understanding how much and where information is nested in sets of data obtained using different parameters and metrics is a typical problem in the analysis of high-dimensional datasets. Approaches based on relative data-driven metrics may be poorly quantitative and difficult to interpret, while accurate methods may be computationally expensive and impractical for large datasets. Here, we report an efficient, quantitative, and transferable method useful to this end, which we call “\textit{Micro-Information Analysis}”. Its foundation is based on the tight relationship that exists between the microscopic features of data distributions and the information they contain. The method performs preliminary unsupervised clustering, capturing microstructural features of data distributions such as the dominant Gaussian microdomains (microstates) present in them, their population, separation, and their residual non-Gaussian fraction. Leveraging the principle of clustering-and-minimization of Shannon entropy, this allows quantifying in \textit{bit} (that is, in absolute metric) the information contained in the microstructure of the data, and how much information is nested in each Gaussian and non-Gaussian microdomain. The output of the analysis is a set of easily interpretable Information-Microstructure plots that allow to readily compare and rank data based on their information content and to graphically map the information breakdown among their various microenvironments in diagrams that are easy to interpret. We prove the efficiency of our method by analyzing and ranking data generated synthetically or from molecular simulations, demonstrating how “\textit{Micro-Information Analysis}” is a physically-grounded quantitative  tool useful for the study of physical systems from their data and for data science in general.
\end{abstract}
%%If you want to cite a paper, do it like this~\cite{Gasparotto2019}. 
\section*{Introduction}

The study of complex systems and phenomena heavily relies on the ability to extract information from data obtained during their observation.\cite{glielmo2021unsupervised,mcgibbon2015mdtraj,michaud2011mdanalysis} Most real-world phenomena cannot be captured by a single series of data or a single dataset; rather, they typically require the acquisition and analysis of large, high-dimensional datasets.\cite{koppen2000curse,verleysen2005curse,pandove2018systematic,steinbach2004challenges} Nonetheless, quantifying the information effectively contained in such datasets and in their individual dimensions may be in many cases far from trivial.\cite{Shannon1948,cover1991elements_infotheory} This requires robust approaches to estimate the information content of a dataset, in a way that is general enough to be transferred to other, and virtually any type of, datasets.\cite{paninski2003estimation,info-imbalance,becchi2025mine,Wild2025-diff-info-imb} Several methods are commonly employed to this end.

One typical approach relies on methods that compress the size of a high-dimensional dataset through dimensionality-reduction techniques. Widely used examples are Principal Component Analysis (PCA)\cite{Jolliffe2016_pca}, time-lagged Independent Component Analysis (tICA),\cite{Molgedey_TiCA,PrezHernndez2013_TiCA,Schwantes2013_tiCA}  which reduce dimensionality on the basis of criteria such as, e.g., the maximum variance, or the slowest-decorrelating linear combinations of the input coordinates (i.e., the slow dynamical degrees of freedom of the system the data refer to), or other ones\cite{Hyvrinen2000,McInnes2018,Tenenbaum2000,Kramer1991}. In these approaches, a large high-dimensional dataset can be projected onto a lower-dimensional space in which the various original dimensions can be then compared and related to each other based on data-driven metrics measuring their mutual similarity/differences.\cite{Martino2025,Gasparotto2019,Capelli2021,Gardin2022}

While effective for a variety of applications, such methods also suffer from some intrinsic limitations, essentially related to the fact that the metrics they use are data-driven (relative) and depend on the supplied data. 
For example, as the number of variables grows, pairwise distances typically tend to lose discriminative power, so that rankings built on very high-dimensional comparisons become increasingly uninformative (the so-called "curse of dimensionality").\cite{koppen2000curse,verleysen2005curse,Lionello2025,becchi2025mine}
Such approaches are useful to assess how similar/different two datasets are by monitoring their relative/mutual distance in the space of the data, but at the same time they do not allow establishing, e.g., which data is more informative.
One method that is particularly useful to this end is, e.g., Information Imbalance\cite{info-imbalance}, a statistical measure quantifying the relative information content of one distance measure with respect to another defined on the same dataset. By comparing distances built from different feature subsets, it can identify which variables are most informative, mutually complementary, or redundant.\cite{Donkor2023,Wild2024,Wild2025-diff-info-imb,Sharma2024} 
In general, while such approaches based on relative metrics are certainly useful to assess, for example, how similar (or similarly informative) different data are to each other, they do not explicitly quantify how much information the data actually contain. Furthermore, the addition of extra data to the comparison also may typically require to re-compute the entire analysis workflow from scratch.  

Other methods aim at directly quantifying the information contained in data. A widely used quantity associated with the information content of data is, e.g., the Shannon entropy\cite{Shannon1948}, which measures the average uncertainty associated with the estimation of specific values of a random variable from a set of measurements. In general, broad data distributions typically exhibit high Shannon entropy, whereas sharply peaked distributions typically have a low Shannon entropy. Another widely used derived quantity is negentropy,\cite{brillouin1953negentropy} defined as the difference between the Shannon entropies of the data distribution and that of a Gaussian distribution with the same mean and variance of the original one.
The advantage of such methods is that they are fast to compute and use an absolute (rather than relative) physical metric to rank the information content of a set of data. One typical limitation of such scalar metrics is that they entail an extreme information compression (from data distributions to one scalar). Consequently, such global (macroscopic) entropy analyses may be ambiguous and not always straightforward to interpret: for example, completely different data distributions may have exactly the same Shannon entropy.

Recently, a Maximum Information Extraction (MInE) method\cite{becchi2025mine} has been reported that, via single-point time-series analysis (e.g., Onion clustering\cite{Becchi2024onion}), allows capturing both the macroscopic and microscopic information contained in time-series data. 
This includes the (macroscopic) information contained in the shape of the data distribution, plus the (microscopic) information contained in the microclusters composing it and in the time-correlations of the data. While such an approach is effective for, e.g., learning the optimal spatiotemporal resolutions (or the best suited descriptors/parameters) for maximum information extraction from noisy time-series data,\cite{becchi2025mine} the considerable computational cost required for such ``deep" time-series analysis makes it prohibitive to apply on large high-dimensional time-series data. 
For the practical purpose of analyzing large high-dimensional datasets of any type (static as well as dynamic data), one would ideally need a method capable of retaining as much as possible such important features in estimating the information content of sets of data, while being at the same time computationally affordable and easy to apply systematically.

Here, we describe a simple, quantitative, and physically-grounded method that we call ``\textit{MicroInformation Analysis}" (MInA).
This is based on the general concept that the information contained in the data is deep-encoded in the microstructural features of their distributions (e.g., the number of Gaussian sub-environments, their separation and sharpness, and the non-Gaussian domains), which can be easily spotted even prior to using elaborate and expensive analysis tools.
By performing preliminary data clustering, our method quantifies the information contained in the data's microstructure, and provides an easy way to map the breakdown of such information content among the various data microdomains in a readily interpretable manner. 
By analyzing different types of examples datasets, either synthetic or obtained from molecular simulations, we demonstrate the efficiency of the method in rapidly ranking up to several hundred dimensions in large datasets based on their information content. All the examples discussed herein underline the tight relationship existing between the microscopic information content of a dataset and the Gaussian and non-Gaussian features of its distribution. We find that our approach provides trends consistent with those obtained with advanced expensive methods (such as MInE), while at the same time allowing to process, screen, and compare a large amount of data impossible to treat with them. This offers a precious tool for the study of complex physical systems from their data, and for data analysis in general.

\section*{Results and discussion}
\subsection{Mapping data microstructures and  micro-information contents}

 \begin{figure}[htbp]
    \centering
    \includegraphics[width=1\textwidth]{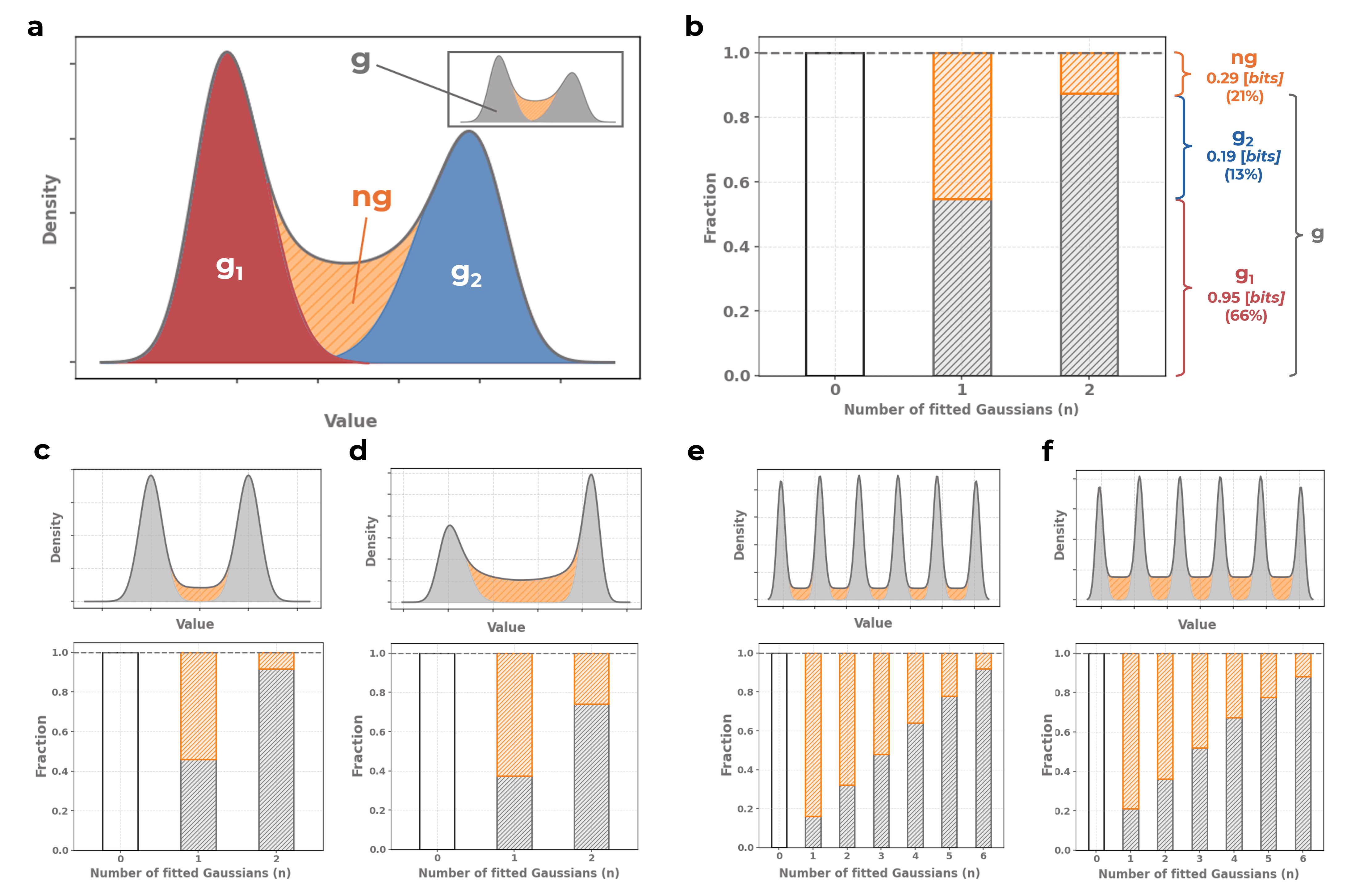}
    \caption{\textbf{Data microstructure diagrams.} (a) Kernel density estimate (KDE) of an example bimodal distribution: the MInA method performs a hierarchical decomposition of the data distribution into Gaussian and non-Gaussian microdomains. The two density peaks are sequentially fitted (proceeding in height descending order) with Gaussian functions, whose areas $g_1$ (red) and $g_2$ (blue) correspond to the fraction of data explained by each domain. The residual region between them (\textit{ng}, orange) is the non-Gaussian fraction of the data that remains once no further peaks can be resolved. The inset shows the Gaussian region of the distribution denoted as \textit{g}. 
    (b) Data microstructure diagrams referring to the data microstructure decomposition of the ditribution of panel (a). The cumulative Gaussian explained fraction $\Sigma\mathrm{\textit{gef}}$ (gray, hatched) and its complementary, the cumulative non-Gaussian fraction $\Sigma\mathrm{\textit{guf}}$ (orange, hatched), increase in stepwise fashion as each successive Gaussian is fitted ($g_1$, then $g_2$), and can be quantified, respectively, as the sum of all gray and orange sub-columns in the diagram. Alongside the diagram, the individual information $\Delta I_n$ (i.e., the information contained in each distinct microenviroments) is reported (on the right of the diagram), both in \textit{bit} and as a percentage of the total information contained in the data microstructure, $\Delta I$. 
    (c-f) Additional synthetic distributions with different Gaussian microdomains and non-Gaussian fractions, captured by the $\Sigma\mathrm{\textit{gef}}$ (gray) and $\Sigma\mathrm{\textit{guf}}$ (orange) quantities, which jointly determine the information contained in the microstructure of the data. For each case, the KDE (top) and the corresponding microstructural plot (bottom) are shown. Panels (c) and (d) compare two distributions with the same number of Gaussian peaks ($n=2$) but different non-Gaussian fractions; panels (e) and (f) compare two distributions with a larger number of peaks ($n=6$), again differing in the prominence of the non-Gaussian regions separating them. Complete data with information contents of the Gaussian and non-Gaussian microdomains are provided in Supplementary Figure~\ref{FigS1}.}
    \label{Fig1}
\end{figure}

The MInA workflow proceeds hierarchically in unveiling the microstructure of the data, and in quantifying and mapping the information contained within it. 
The approach is general, and the data may be of any type: a distribution of arbitrary shape in which one or more density peaks can be identified. A few representative examples are reported in Figure \ref{Fig1}. 

Our approach builds on the concept that the presence of multiple clearly distinguishable density peaks in the data distribution carries \textit{per se} relevant physical information, each peak identifying a different and highly populated microstate in the system the data refer to, which we assume to be Gaussian-like. In this way, the number of different Gaussian microdomains/microstates that can be distinguished, their population, prominence, sharpness, and separation, are all key microscopic features of the data distribution that can be directly related to the information contained in the data/system's microstructure. 
Notably, under this premise, the basins between the different peaks also contain relevant information, identifying, for example, the presence of transition states/patterns between the different microstates. 

From the algorithmic point of view, a MInA analysis proceeds in this way: for each example dataset analyzed herein, (i) Gaussian microstates (\textit{$g_n$}) are hierarchically identified through microscopic-level KDE peak detection (from the highest to the lowest one) and local Gaussian fitting; (ii) each data point is assigned to its respective Gaussian environment of belonging (based on a highest likelihood principle); (iii) all remaining data points (not belonging to microenvironments with detectable density peaks)  are assigned to a residual non-Gaussian cluster (\textit{$ng$}). 

As a first representative example, we use the distribution of Figure \ref{Fig1}a. The MInA flow described above hierarchically discovers a first density peak onto which a first Gaussian is fitted, \textit{$g_1$} (in red), then a second one, \textit{$g_2$} (in blue), and a residual non-gaussian domain, \textit{$ng$} (dashed orange).
At the same time, starting from the first fitted Gaussian, the method starts drafting the ``data microstructure diagram" of Figure \ref{Fig1}b. 
Note that, before starting, no Gaussian has yet been fitted, and the whole dataset is still unexplained in terms of Gaussian and non-Gaussian content (Figure \ref{Fig1}b: initial empty bar). When the first Gaussian (Figure \ref{Fig1}a, in red) is fitted, the method computes the fraction of points belonging to \textit{$g_1$}, that is, the fraction of the data explained by this first Gaussian microdomain. In the specific example of Figure \ref{Fig1}a, \textit{$g_1$} contains $\sim 51\%$ of the data, so that, after its detection, the Gaussian explained fraction increases to $\sim 0.51$ in the data microstructure diagram of Figure \ref{Fig1}b (gray bar at $n=1$). 
The method then identifies a second Gaussian microcluster, \textit{$g_2$}, containing additional $\sim 35\%$ of data points, so that the total Gaussian explained fraction increases to $\sim0.86$ after its detection (Figure \ref{Fig1}b: gray bar at $n=2$). 
In this example distribution, no further density maxima can be detected. Based on the MInA method, the total Gaussian fraction of this dataset, in terms of data content, is $g=g_1+g_2\sim86\%$ (subdivided in the two Gaussian microenvironments, \textit{$g_1$} and \textit{$g_2$}), while the complementary non-Gaussian fraction is $ng\sim14\%$ (the fraction of data belonging to the $ng$ microdomain). 

From such a data microstructure diagram (Figure \ref{Fig1}b), it is interesting to compute the cumulative Gaussian explained fraction $\Sigma{\mathrm{\textit{gef}}}$ (sum of all dashed gray bars), and its counterpart, the cumulative Gaussian unexplained fraction $\Sigma{\mathrm{\textit{guf}}}$ (sum of all dashed orange bars). In particular, $\Sigma{\mathrm{\textit{gef}}}$  captures not only the total Gaussian fraction of the data, but also how many Gaussian microenvironments are necessary to reach it, while its complementary, $\Sigma{\mathrm{\textit{guf}}}$, contains information on how much of the data remains non-gaussian despite the discovery of the $g_n$ Gaussian microdomains. 

These parameters, readily available from the microscopic clustering and data microstructure diagrams performed by MInA, implicitly capture (being dependent on) how many Gaussian microdomains are present in the data, how diverse they are in terms of sharpness/broadness, populations, how well-separated they are, and how large the non-Gaussian part is compared to the Gaussian one. This is relevant in that, for example, data distributions containing two putative Gaussian microenvironments, $g_1$ and $g_2$, may have the same total Gaussian fraction $g$ while having different data microstructure diagrams, and different $\Sigma{\mathrm{\textit{gef}}}$ and $\Sigma{\mathrm{\textit{guf}}}$. In fact, the latter depend on characteristic microscopic features of the data distribution, such as the diversity or similarity between their Gaussian microdomains, etc. A data distribution having a very large, dominant Gaussian microcluster $g_1$ and a tiny secondary one $g_2$ will have, for example, a larger $\Sigma{\mathrm{\textit{gef}}}$ and a smaller $\Sigma{\mathrm{\textit{guf}}}$ (a larger gray area and a smaller orange one in the diagrams of Figure \ref{Fig1}) compared to a case where $g_1 \approx g_2$ in terms of population, even if the total $g$ is the same.    
Note also that the weights, in terms of population, of all Gaussian and non-Gaussian microclusters composing the data do not correspond necessarily to their information content (which depend also on the shape of the microclusters).

In fact, in parallel, a similar breakdown of the information content is performed in the MInA workflow by estimating the information contained in the microstructure of the data, and its partition among the various microclusters. These quantities can be readily calculated by leveraging the principle of maximum information extraction via clustering-and-minimization of Shannon entropy\cite{becchi2025mine}, and by considering the statistical population weights of the various microclusters.
The central idea behind our approach is that decomposing the data into the Gaussian and non-Gaussian microdomains effectively distinguishable in the data implicitly provides an information gain, effectively revealing the information contained in the data's microstructure.

The Shannon entropy\cite{Shannon1948} of an observable $x$, which takes values in a set $\chi$ with probability distribution $p(x)$, is defined as:

\begin{equation}
    H(x) = -\sum_{x \in \chi} p(x) \log_2 p(x)    
    \label{eq:eq1}
\end{equation}

Our method first computes the Shannon entropy of the entire data distribution, which we define as $H_0$.
Then, it calculates the Shannon entropies of the various Gaussian ($H_{g_1}$, $H_{g_2}$, ... , $H_{g_n}$) and non-Gaussian ($H_{ng}$) microclusters present in the data. 
For a dataset composed of $n$ clusters with probability distributions $p_n(x)$, the Shannon entropy of the clustered dataset can be calculated as the weighted sum of the Shannon entropies of the microclusters\cite{becchi2025mine}:

\begin{equation}
    H_{clust}(x) = \sum_{i=1}^{n} w_nH_n    
    \label{eq:eq2}
\end{equation}

where $w_1,..., w_{n}$ are the fraction of data belonging to the various clusters (i.e., their populations).

By concavity, the subdivision of the data into microclusters reduces entropy\cite{becchi2025mine}: that is, the Shannon entropy of the whole distribution is always higher than, or equal to, the weighted sum of the Shannon entropies of the microclusters (the equality holding in the limit case where no distinct microclusters are present in the data). 
By comparing $H_0$ and $H_{clust}$, it is thus possible to calculate the information gain ($\Delta I$) as:

\begin{equation}
    \Delta I = H_{0}-H_{clust} \geq 0    
    \label{eq:eq3}
\end{equation}

In general, the information gain $\Delta I$ obtained following data clustering corresponds to the information contained in the (discovered) microstructure of the data that, in our workflow, is measured in $[bit]$. 

The information contained in each of the $n$ microclusters present in the data depends on its relative statistical information weight, and is readily quantifiable as:

\begin{equation}
    \Delta I_n = w_n(H_0-H_{n})   
    \label{eq:eq4}
\end{equation}

In the specific case of the example distribution of Figure \ref{Fig1}a), the Shannon entropy of the distribution is $H_0=4.85 ~[bit]$.
Considering the percentage populations of the two Gaussian and non-Gaussian microclusters reported above, their individual entropies are readily calculated in MInA, and $H_{clust}$, as defined in Eq. \ref{eq:eq2}, can be calculated as: 

\begin{equation}
   \begin{split}
    H_{clust} &=w_{g1}H_{g1}+w_{g2}H_{g2}+w_{ng}H_{ng}\\
    &=0.51 \cdot 2.99~[bit]+0.35 \cdot 4.31~[bit]+0.14 \cdot 2.78~[bit]\\
    &=1.52~[bit]+1.51~[bit]+0.39~[bit]= 3.42~[bit]    
    \label{eq:eq5}
   \end{split}
\end{equation}

and, based on Eq. \ref{eq:eq3}, the information contained in the microstructure of the distribution equals to: $\Delta I = 1.43~[bit]$.

Quantitatively, this is a first relevant insight. By definition, a dataset containing two possible microstates with equal statistical weights has a Shannon entropy of $H=1~[bit]$ ($=-\log_2(0.5)$). This means that the data distribution of Figure \ref{Fig1}a contains more information than that related to the $g1$ and $g2$ Gaussian microdomains. Actually, this suggests that in this dataset there is (at least) one additional microenvironment, which information content is comparable to those of the other two main Gaussian microstates (considered that the Shannon entropy of an ideal equiprobable three-states system is $H=-\log_2(0.3\overline{3})=1.585~[bit]$). 
Based on Eq. \ref{eq:eq4}, the information contained in each of the three microclusters is: $\Delta I_{g1} = 0.95~[bit]$, $\Delta I_{g2} = 0.19~[bit]$, and $\Delta I_{ng} = 0.29~[bit]$. These values are readily reported alongside the data microstructure diagram of Figure \ref{Fig1}b, and allow one to easily appreciate the imbalance between the population and the information content of the various microenvironments present in the data.

For example, in Figure \ref{Fig1}b it is interesting to note how, in the example distribution of Figure \ref{Fig1}a, the first Gaussian microstate $g_1$ contains $\sim51 \%$ of the data points but $0.95~[bit]$ of information, which corresponds to as much as $\sim 66 \%$ of the total information contained in the data microstructure. The data microstructure diagram of Figure \ref{Fig1}b also shows how, interestingly, the non-Gaussian microcluster $ng$ contains more information than the second Gaussian microstate $g_2$ ($0.29~[bit]$ vs. $0.19~[bit]$, respectively) despite being less populated ($\sim14 \%$ vs. $\sim35 \%$, respectively). This is due to the fact that its entropy is relatively low $H_{ng}=2.78~[bit]$, similar to that of $g1$, and way lower than that of $g2$.
Considering the fact that the $ng$ microcluster does not correspond to a Gaussian peak and is thus related to a non-Gaussian transition microstate, this indicates that such a transition microdomain contains a relevant pattern, and relevant information that are just hidden by the height of the $g1$ and $g2$ microclusters.
This is demonstrated in Supplementary Figure~\ref{FigS2}: a detection-classification-deletion flow similar to that used by, e.g., Onion clustering\cite{Becchi2024onion}, where after their detection, $g1$ and $g2$ are sequentially deleted from the data distribution, shows that the remaining $ng$ environment has $per se$ a clear sharp structure (with low Shannon entropy).  

Such non-obvious insight is relevant for practical data analysis purposes. In fact, in the hypothetical case in which such synthetic data distribution referred to time-series data, this would suggest that it is worth analyzing it with single-point time-series analysis approaches (e.g., with Onion clustering\cite{Becchi2024onion}) capable of extracting information related to the transitions and time-correlations between the data-points. This information is useful especially when dealing with large high-dimensional datasets, where applying accurate but computationally heavy analyses -- such as performing an Onion\cite{Becchi2024onion} clustering analysis for each individual dimension -- would be impractical. There, MInA can be extremely useful, quickly unveiling the subset of data worth a second round of analysis with dedicated advanced approaches. 
While this will be discussed in detail in the next sections below, we anticipate that, in general, this is the case for all datasets in which the information content of $ng$ and the $\Sigma\mathrm{\textit{guf}}$ are relatively large. Conversely, the added value of such additional analyses will become more and more negligible in data distributions dominated by Gaussian microstates (where $ng\sim 0$, in terms of population and information content).

The same MInA analysis can be applied to virtually any type of data distribution. Figure \ref{Fig1}c-f reports other example synthetic datasets differing in the number and arrangement of Gaussian peaks, as well as in the non-Gaussian fraction they contain. 
In general, the higher is the number of microclusters that can be discovered in the data distribution, the higher is the information contained in their microstructure, $\Delta I$. For example, $\Delta I$ equals to $1.32~[bit]$ and $1.53~[bit]$ in the two bi-peaked distributions of Figures \ref{Fig1}c and \ref{Fig1}d, while it is has high as $2.73~[bit]$ and $2.33~[bit]$ in the six-peaked distributions of Figures \ref{Fig1}e and \ref{Fig1}f. 
Aside from such a clear relationship between the number of main Gaussian peaks, the structure of the data, and their information content, these data's microstructure diagrams capture also other microscopic-level features that are relevant for evaluating the information contents.
Comparing the distributions of Figures \ref{Fig1}c and \ref{Fig1}d demonstrates how the number of Gaussian peaks alone is not sufficient to capture their differences. Changing the sharpness, broadness, separation, and population of the Gaussian microclusters has a clear impact on the non-Gaussian fraction of the data (in orange, larger in Figure \ref{Fig1}d than in \ref{Fig1}c). Interestingly, even though the Gaussian peaks are less high and less well-defined in the KDE of Figure \ref{Fig1}d,  its microstructure nonetheless contains more information than that of Figure \ref{Fig1}c , which suggests that the $ng$ microdomain, and its relationship with the $g$ ones, plays a prominent role.

The relationship between the Gaussian and non-Gaussian fractions of the data and the information that these contain is captured in the data's microstructure diagrams by the $\Sigma\mathrm{\textit{gef}}$ and $\Sigma\mathrm{\textit{guf}}$ parameters (the gray vs. orange areas). By comparing the data of Figure \ref{Fig1}c-f, and the $\Delta I$ values provided above, it is clear that, at a constant number of peaks, data distributions with a more pronounced $ng$ fraction and a larger $\Sigma{\mathrm{\textit{guf}}}$ carry more information.
When a number of Gaussians is fitted, the $\Sigma{\mathrm{\textit{guf}}}$ remains relatively high (and the $\Sigma{\mathrm{\textit{gef}}}$ low), the physical interpretation is that there is more information nested in the data than that contained within those Gaussian microenvironments alone.
 
As an aside comment, it is worth noting that such unsupervised clustering of a signal, or of a dataset in general, into Gaussian and non-Gaussian microclusters differs, both conceptually and technically, from what done, for example, by a Gaussian Mixture Model (GMM)\cite{Reynolds2009}. A GMM is a probabilistic ``numerical method" that fits the entire data with a certain number of Gaussians. Below the minimum number of Gaussians necessary to obtain an optimal fit and recover the features of the data distribution, the individual Gaussians that are deposited do not carry a direct physical meaning (they do not capture, e.g., a subset of the physical microstates of the system under study, and not others), but merely a numerical one.  
One important difference is that, in placing its Gaussians, a GMM fits all the data macroscopically and in one shot, whereas our method proceeds in an exquisitely microscopic fashion.
In particular, our method does not aim to cluster all data in Gaussian microclusters (as GMM does), but only in those accompanied by clear density peaks, which can thus be related to distinct, statistically relevant physical microclusters in the system under study (while the fraction of data not belonging to them is classified as part of the residual non-Gaussian domain). In that, the clustering performed by the MInA method is exquisitely microscopic and physically-grounded (rather than numerical), uncovering in hierarchical fashion all Gaussian and non-Gaussian microstates present in the data, which contain different physical information.

\begin{figure}[htbp]
    \centering
    \includegraphics[width=1\textwidth]{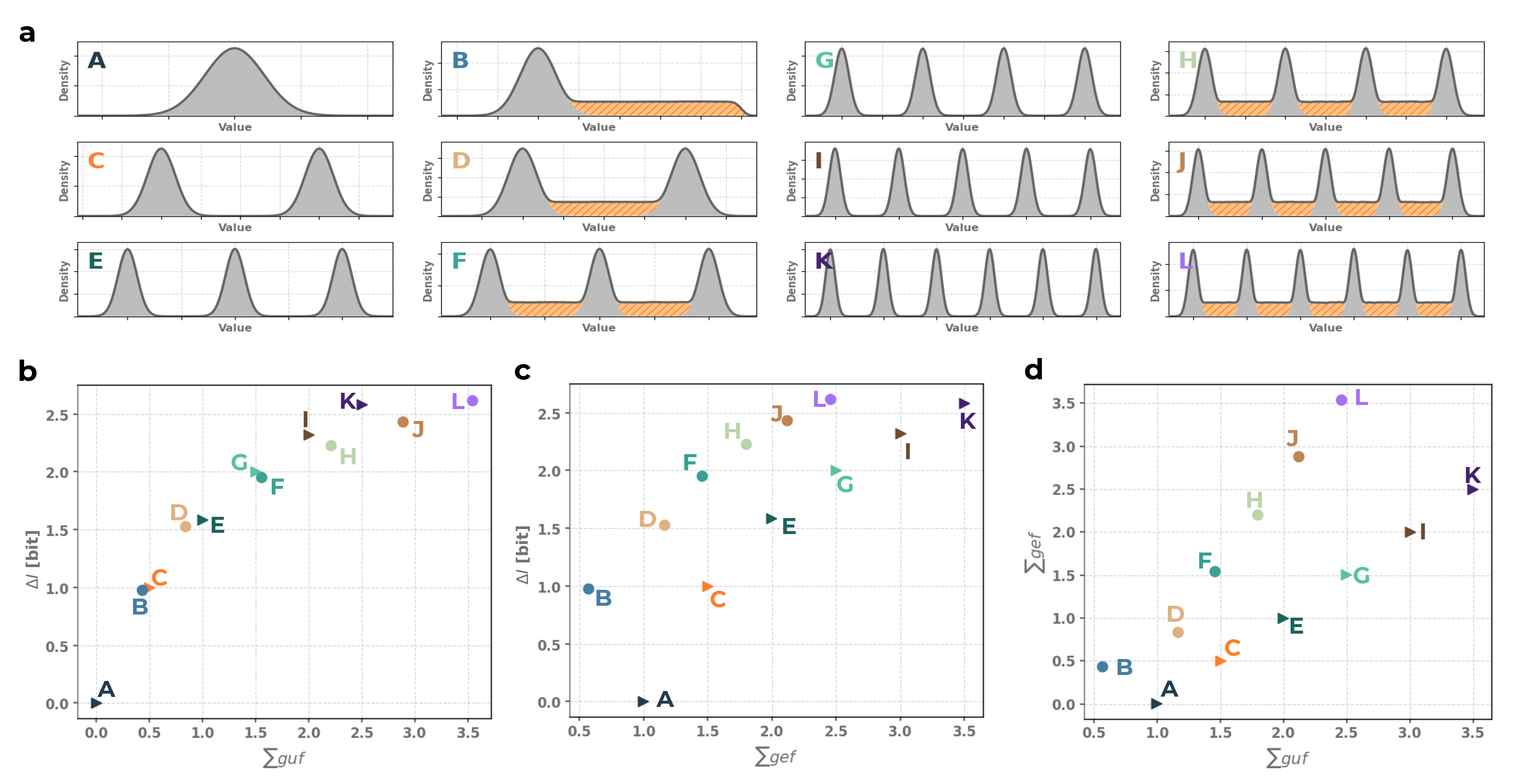}
    \caption{\textbf{Information-Microstructure plots.} (a) Example KDEs of twelve synthetic distributions (\textit{A}-\textit{L}), differing in the number of Gaussian peaks (1 to 6) and in the size of the non-Gaussian fraction (orange hatched) separating them, at fixed peak number. (b) Information contained in the data microstructure, $\Delta I$ (in $[bit]$), as a function of the cumulative Gaussian unexplained fraction $\Sigma\mathrm{\textit{guf}}$ for all twelve datasets. (c) $\Delta I$, in (in $[bit]$), as a function of the cumulative Gaussian explained fraction $\Sigma\mathrm{\textit{guf}}$. (d)$\Sigma\mathrm{\textit{guf}}$ as a function of $\Sigma\mathrm{\textit{gef}}$ for the same datasets (the data move bottom-left to top-right as a function of the number of Gaussian microenvironments that are detected in the data).}
    \label{Fig2}
\end{figure}

In Figure \ref{Fig2}a, we show additional example data distributions (KDEs) differing in the number of Gaussian peaks and in the presence of a non-Gaussian fraction (\textit{ng}, colored in orange). For each KDE, we computed the associated microstructure diagram, which allows examining the trends encoded in the different microstructural variables it captures.
It is interesting to plot key microstructural features, such as $\Sigma\mathrm{\textit{guf}}$ and $\Sigma\mathrm{\textit{gef}}$, against the information contained in the data's microstructure, $\Delta I$. The resulting trends are reported in Figure \ref{Fig2}b-d, in the form of information-microstructure plots.

A striking, direct relationship emerges between $\Sigma{\mathrm{\textit{guf}}}$ and $\Delta I$ (Figure \ref{Fig2}b). The fact that such a direct relationship is typically non-linear (see also next sections below) indicate that $\Sigma{\mathrm{\textit{guf}}}$ discriminates better between the most informative datasets, while the $\Delta I$ parameters is more efficient in discriminating between the least informative ones.
This suggests that, as a single parameter/metric, $\Sigma{\mathrm{\textit{guf}}}$ is more effective for ranking such data based on their information content by implicitly capturing (as said above) much of the relevant information contained in the data's microstructure in a single parameter. In fact, in the data microstructure diagrams, $\Sigma{\mathrm{\textit{guf}}}$ is the cumulative sum of all orange bars (see Figures \ref{Fig1}b-f), and it depends directly on the number of Gaussian microclusters that have been detected in the data (the $x$-axis of the diagrams), on their populations, as well as on the residual non-Gaussian fraction not explained by them.

For example, the information-microstructure plot of Figure \ref{Fig2}b clearly shows how the datasets with six Gaussian microenvironments are the most informative ones, and evidently more than the data distributions displaying four, three, two or one single Gaussian peaks. 
In particular, the (A) dataset, which is nearly one Gaussian, has, accordingly to the MInA method, negligible information gain obtained by unveiling the data microstructure: $\Delta I \sim 0$. On the other hand, (B), which is like a skewed Gaussian, has more information stored into its microstructure ($\Delta I \sim 0.8~[bit]$), due to the fact that it has essentially two microstates distinguishable in it: one dominant Gaussian micrcluster (Figure \ref{Fig2}b: in gray) plus a non-negligible residual non-Gaussian one (dashed orange).

The results obtained for data (C) and (D) are also interesting, being their $\Delta I \sim 1.40-1.55~ [bit]$ (not far from that expected for an ideal system with three equiprobable microstates, $\sim 1.585~ [bit]$). As seen above, this indicates the relevant information content of the non-Gaussian $ng$ domain in such two-Gaussian-peaked distributions. Similar trends are evident, e.g., in the (H) and (L) datasets, where the information content exceeds those expected for ideal four- or six-microstates systems (in theory, respectively $2$ and $2.59~[bit]$).
Based on what said above, (H) should contain more information nested within it than (G), having a higher non-Gaussian fraction together with the four Gaussian peaks, and the same is true for (L) vs. (K). This principled theoretical ranking is exactly reflected by $\Sigma{\mathrm{\textit{guf}}}$, but less by $\Delta I$.
The fact that the $\Delta I-\Sigma{\mathrm{\textit{guf}}}$ trend is direct but not linear thus reinforces the concept proposed above that $\Sigma{\mathrm{\textit{guf}}}$ offers a very efficient metric for ranking and discriminating between different datasets (and especially between the most informative ones). Thiswill be further elaborated in the following sections, and will become strikingly clear when dealing with larger, high-dimensional datasets.  

Comparing Figure \ref{Fig2}b with Figure \ref{Fig2}c shows a different trend of $\Sigma{\mathrm{\textit{gef}}}$ against $\Delta I$. Here, comparing (A) vs. (B), (C) vs. (D), etc., shows how $\Sigma{\mathrm{\textit{gef}}}$ caèpture the effect of growing numbers of Gaussian microclusters in the data, but not the information content of the $ng$ microcluster. 
In particular, while by their own definitions $\Sigma{\mathrm{\textit{guf}}}$ and $\Sigma{\mathrm{\textit{gef}}}$ are both directly proportional to the number of Gaussian microclusters in the data, at the same time, they are inversely related to each other for data distributions containing the same number of peaks while in Figure \ref{Fig2}d this will be even clearer later on, when dealing higher-dimensional datasets), they are represented by the gray and orange complementary areas in the microstructure plots shown in Figure \ref{Fig1}. 

In general, the MInA workflow thus performs a preliminary microscopic clustering of the data, and quantifies and represents the information contained in their microstructure. The analysis presented in Figures \ref{Fig1} and \ref{Fig2} shows that the proposed method is simple, flexible, efficient, and, importantly, physically interpretable by construction. This offers an efficient method to rank data of different nature according to the information contained in their microstructure through an objective, physically interpretable, and transferable metric.

\subsection{Micro-Information Analysis of heuristic molecular dynamics data}
\begin{figure}[htbp]
    \centering
    \includegraphics[width=0.9\textwidth]{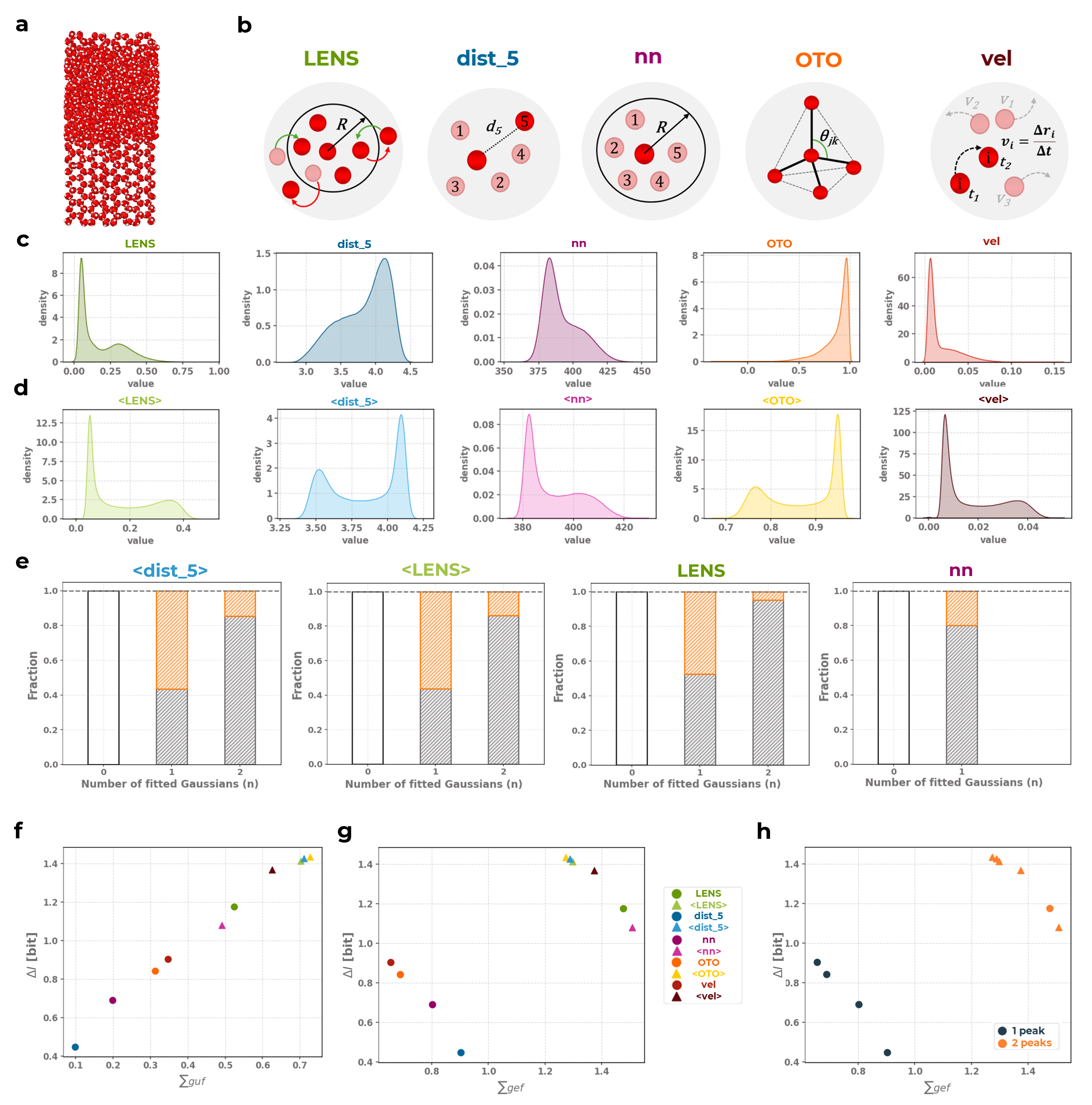}
    \caption{\textbf{Micro-Information Analysis of example heuristic molecular dynamics datasets.} (a) MD snapshot of the simulated aqueous system of water at solid-liquid coexistence (see Methods section for technical details). (b) Schematic representation of the five example descriptors used herein to analyze the MD trajectories. Left to right: Local Environment and Neighbor Shuffling ($LENS$), distance to the fifth neighbor ($dist\_5$), number of neighbors ($nn$), orientational tetrahedral order ($OTO$), and velocity ($vel$). (c) KDE of the raw data obtained by analyzing the MD trajectories with the different descriptors ($D$). Among these raw datasets, the $LENS$ one (in green) is the unique one with a KDE clearly showing two density peaks (while all others have just one). (d) KDE of the corresponding denoised data obtained with the various descriptors ($\langle D \rangle$), obtained by spatial averaging (see methods for technical details).\cite{Donkor2024spatavg,Martino2025} After local noise reduction via spatial averaging, all denoised KDE show the density peaks (referring to the ice and liquid water phases in the system).\cite{Martino2025}  (e) Data microstructure diagrams for four representative datasets obtained with the $\langle dist\_5 \rangle$, $\langle LENS \rangle$, $LENS$, and $nn$ descriptors. For each descriptor, the cumulative Gaussian explained fraction $\Sigma\mathrm{\textit{gef}}$ (gray) and non-Gaussian fraction $\Sigma\mathrm{\textit{guf}}$ (orange) are shown as a function of the number of fitted Gaussians, $n$ (among these, the $nn$ dataset is the only one with just one single Gaussian peak/microcluster - see also the KDE of panel (c), in violet). (f-h) Information-microstructure plots reporting $\Delta I$ as a function of the cumulative Gaussian unexplained fraction $\Sigma\mathrm{\textit{guf}}$ (f) or $\Sigma\mathrm{\textit{gef}}$ (g,h), for the raw (circle) and denoised (triangle) datasets obtained with the various descriptors. In (h), the data points are colored based on the number of Gaussian peaks/microclusters detected in the KDEs they refer to, showing a neat separation between single- (black) and two-peaked datasets (orange) on the $\Sigma\mathrm{\textit{gef}}$-$\Delta I$ plane. }
    \label{Fig3}
\end{figure}

To test the practical utility of the method, we applied the MInA approach on data extracted from an example molecular dynamics (MD) simulation trajectory of 2048 TIP4P/ICE molecules, initially $50\%$ in ice and $50\%$ in liquid water phases, which are simulated at their coexisting temperature (Figure \ref{Fig3}a, complete simulation details are available in the Methods section).\cite{Caruso2023,Crippa2023,Becchi2024onion,Martino2025,Doria2025,Lionello2025,Caruso2025,becchi2025mine,Doria_timederiv} 

One typical first step to study the physics of a system such as this one is to select a well-suited descriptor to extract information from the trajectories of its constitutive units (molecules). Nonetheless knowing \textit{a priori} which descriptor is best for extracting information from MD trajectories and, in general,  even comparing descriptors in this respect, is often far from trivial. 
Here we chose a few descriptors that we use to analyze the MD trajectories of this example system under study. The descriptors selected for this specific test case span a spectrum, from the most abstract and general to those tailored \textit{ad hoc} for the study of aqueous systems and for detecting, e.g., different types of domains or phases present in the system (see their schematic representation in Figure \ref{Fig3}b).\cite{Martino2025,Crippa2023,Becchi2024onion,Doria2025} 
For example, the Local Environment and Neighbor Shuffling (\textit{LENS}) is an abstract descriptor that measures the rate at which the neighbors of each water molecule in the system reshuffle over time.\cite{Crippa2023} Despite its generality and abstractness, this descriptor was found very efficient to discriminate, e.g., solid from liquid phases and the molecular exchange transitions between them.\cite{Crippa2023,Becchi2024onion,Caruso2025,becchi2025mine,Doria_timederiv,Doria2025} The distance to the fifth nearest neighbor (\textit{$dist\_5$}) and the Orientational Tetrahedral Order (\textit{OTO}) are two widely used descriptors for studying water systems (leveraging prior knowledge that ice possesses a well defined tetrahedral local symmetry that is absent in liquid water). The number of neighbors (\textit{nn}) is a very general descriptor that counts the neighboring molecules within a cutoff radius from each molecule in the system. In addition, the velocity of each water molecule (\textit{vel}), computed as its displacement over a fixed time interval, is another general descriptor.  
Notably, \textit{nn}, \textit{OTO}, and \textit{$dist\_5$} are $f(t)$, while \textit{LENS} and \textit{vel} are $f(\Delta t)$. One timeseries dataset was computed for each different from the MD trajectory ($50~ns$ sampled every $40~ps$) under study. 

Since each time-series is characterized by an intrinsic signal-to-noise ratio (that depends on the descriptor and on the internal physics of the system under study\cite{Martino2025}), in this analysis we considered data distributions obtained both from the raw (Figure \ref{Fig3}c) and the denoised data obtained with the various descriptors (Figure \ref{Fig3}d). For each descriptor dataset $D$, the corresponding denoised data distributions is denoted as $\langle D \rangle$ (Figure \ref{Fig3}). In particular, denoising was performed with the spatial-averaging approach \cite{Donkor2024spatavg, Martino2025}, in which the value of a descriptor for each molecule at each frame was replaced with the average of its value and those of the neighbor molecules within a certain cutoff radius (reducing the effect of local noise in the data). 

In principle, a KDE distribution of a dataset allowing to easily describe a two-state system, here ice and liquid water, should be characterized by two distinct peaks (one for each state) separated by a transition region representing the interface between them.\cite{Martino2025,Lionello2025} 
In this case, among the raw descriptors of Figure \ref{Fig3}c, only LENS is able to resolve the two states into separate density peaks, while the others are characterized by a single peak (ice) accompanied by a more or less pronounced shoulder or tail. This confirms that the LENS descriptor is efficient in distinguishing the solid from the liquid phase of water compared to the other descriptors tested herein.\cite{Martino2025}
Denoising the signals, on the other hand, suppresses single-molecule noise and better emphasizes structural information from more distant neighbor molecules, which, in general, reshapes the distributions and sharpens the detected Gaussian peaks.\cite{Martino2025,Lionello2025}
As shown in Figure \ref{Fig3}d, all denoised data distributions exhibit two peaks, allowing ice and water to be readily distinguished and classified with simple clustering methods (for details on the descriptors and denoising we refer the reader to the Method section and to a dedicated paper on this topic by \citet{Martino2025}).

For each raw ($D$) and denoised ($\langle D \rangle$) descriptor data distributions, MInA resolved the corresponding microstructural diagram, $\Sigma\mathrm{\textit{guf}}$ and $\Sigma\mathrm{\textit{gef}}$. 
Figure \ref{Fig3}e shows the data microstructure diagrams of four representative cases (the remaining diagrams are reported in Supplementary Figure~\ref{FigS3}). 

Figure \ref{Fig3}f-h report the same information-microstructure plots as in Figure \ref{Fig2}b-d. $\Sigma\mathrm{\textit{guf}}$ and $\Delta I$ (Figure \ref{Fig3}f) exhibit the same direct trend as in Figure \ref{Fig2}b. Based on these data, again the descriptors can be grouped into two families based on the number of Gaussian microenvironments detected by MInA.  The first, identified by the data distributions located approximately in the region with $\Delta I <1.0 $ [\textit{bit}]  and  $\Sigma\mathrm{\textit{guf}} < 0.4$, includes the less informative descriptors, i.e., those showing one single clearly detectable Gaussian peak. The second family includes highly informative descriptors, which distributions have two detectable Gaussian microenvironments, with $\Delta I > 1.0$ [\textit{bit}] and $\Sigma\mathrm{\textit{guf}} > 0.4$. In this plot, the ordering of the most informative descriptors along the $\Sigma\mathrm{\textit{guf}}$ axis also closely follows the ranking and their ability resolve the water, ice, and ice-liquid interface environments (as a non-Gaussian microcluster), as recently reported and discussed in detail in \citet{Martino2025} using (heavier) Onion clustering timeseries analyses. Figure \ref{Fig3}g shows a similar behavior for $\Delta I$ as a function of $\Sigma\mathrm{\textit{gef}}$, where the two families are clearly separated. In the $\Sigma\mathrm{\textit{gef}}$-$\Delta I$ information-microstructure plot, the most informative datasets are those in the top-right area of the plane (orange points in Figure \ref{Fig3}h), and, more precisely, in its top-left region (in this case, $\langle dist\_5 \rangle$, $\langle OTO \rangle$, and $\langle LENS \rangle$).
These are the same data points which are found on the top-right corner of Figure \ref{Fig3}f (being $\Sigma\mathrm{\textit{gef}}$ and $\Sigma\mathrm{\textit{guf}}$ inversely proportional for the same number of detected peaks).

In general, data denoising sharpens, or even increases the number of the detected Gaussian peaks. Also, it increases their separation, which typically widens the non-Gaussian region lying between them (Figure \ref{Fig3}c-d). As a result, a larger fraction of the data falls in the non-Gaussian part, and, consequently, $\Sigma\mathrm{\textit{guf}}$ increases, and $\Delta I$ generally increases substantially (Figure \ref{Fig3}e-h). This is precisely what the MInA method captures, confirming that a substantial fraction of the useful information resides in the non-Gaussian part of the data. 
This is a rather general effect seen in all the different descriptors datasets. In the case of LENS, the effect of denoising is much smaller, confirming how LENS is, by construction, less noisy and efficient even in raw version (two peaks even in its raw KDE).\cite{Martino2025,becchi2025mine} 

The results in Figure \ref{Fig3} demonstrate how MInA can perform a detailed comparison of the information content of different types of data, leveraging an objective quantitative metrics ($\Delta I$, $\Sigma\mathrm{\textit{guf}}$, $\Sigma\mathrm{\textit{gef}}$), and despite the fact that the individual native metric of each descriptor is \textit{per se} different. 
Notably, the results in Figure \ref{Fig3} are consistent with those obtained in precedence using a high-dimensional, data-driven, similarity-based metric (PCA-based) on the same dataset\cite{Martino2025}. However, in this case, this is possible without building a high-dimensional space containing all data and relying on a data-driven metric, but using an objective, transferable, and physically interpretable metric. The utility of the method, and the trends observed here become particularly clear when applying MInA to larger high-dimensional datasets as in the next sections.

\subsection{Information search and ranking in high-dimensional datasets}

\begin{figure}[htbp]
    \centering
    \includegraphics[width=1\textwidth]{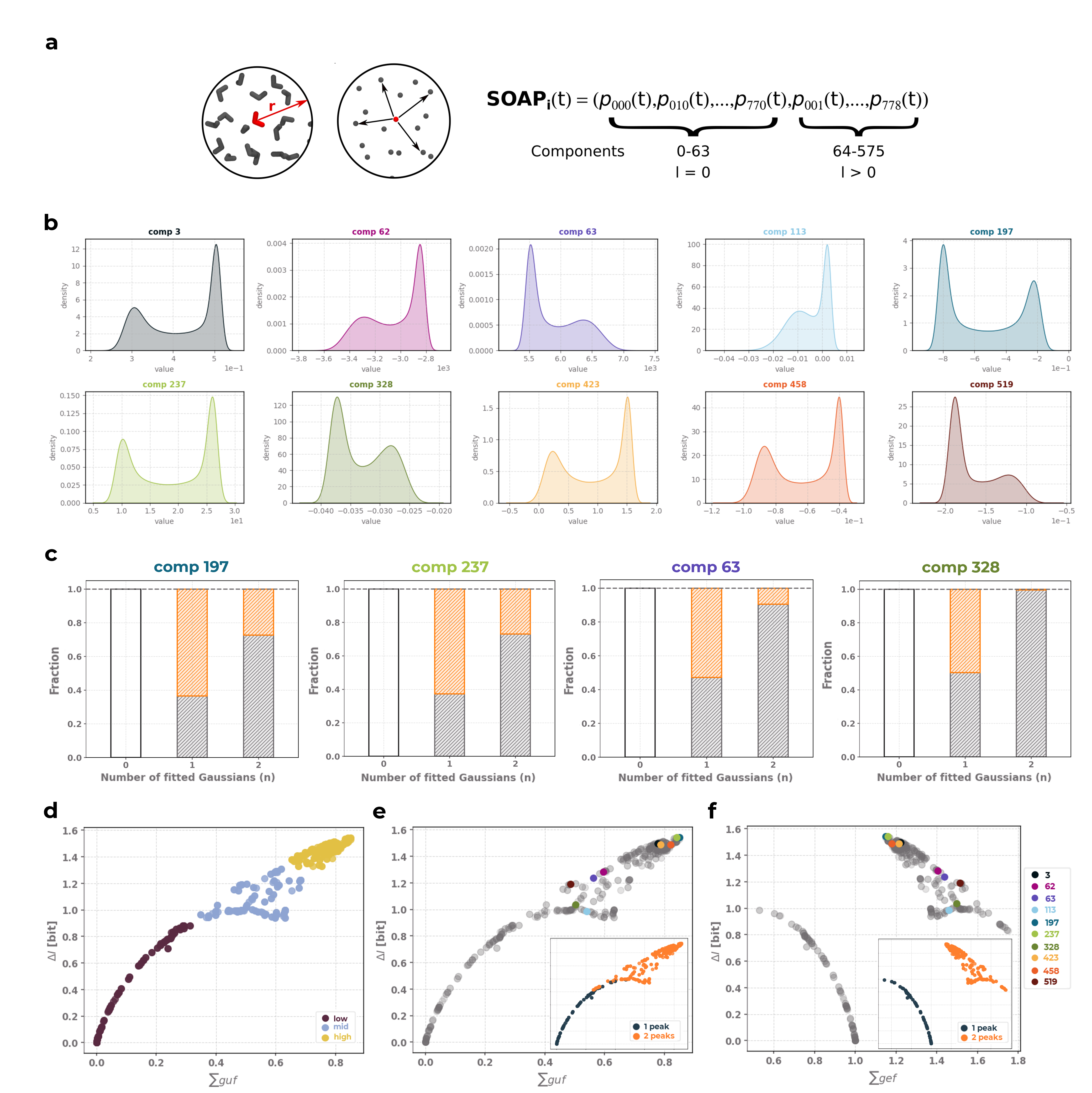}
    \caption{\textbf{Analyzing an example high-dimensional $\langle SOAP \rangle$ dataset with MInA.} (a) Scheme of the SOAP descriptor. SOAP provides a high-dimensional description of the order/disorder in the displacement of the neighbors around each molecule in the system. In this case, SOAP specra are calculated for each of the 2048 water molecules from the MD trajectories described in the previous section (Figure \ref{Fig3}a). At each (1250) MD time-frame, 576-dimensional SOAP spectra are calculated for each of the 2048 water molecules in the system (and further denoised via spatial averaging to reduce local noise as described above).\cite{Lionello2025,Martino2025,Donkor2024spatavg} The dataset studied herein thus contains $2.56 \times 10^6$ $\langle SOAP \rangle$ spectra which are 576-dimensional (see Methods for further details). Each of the 576 $\langle SOAP \rangle$ components contains data with own distributions, which are analyzed with MInA. (b) KDEs of the data of ten representative $\langle SOAP \rangle$ components (component number on the top of each plot). (c) Data microstructure plot for four example components ($\langle 197 \rangle$, $\langle 237 \rangle$, $\langle 63 \rangle$, $\langle 328 \rangle$). (d-f) Information-microstructure plots showing $\Delta I$ as a function of $\Sigma\mathrm{\textit{guf}}$ (d) for all 576 $\langle SOAP \rangle$ components, colored based on their information content (low, mid, high), (e) with the 10 components highlighted in (b) in colors against the full set of components in gray (inset: components colored by the number of detected Gaussian peaks). (f) $\Sigma\mathrm{\textit{gef}}$-$\Delta I$ plots colored in the same way as in (e). Analogous MInA analyses for the raw (non-denoised) SOAP dataset are available in Supporting Figure~\ref{FigS5}.}
    \label{Fig4}
\end{figure}

High-dimensional analyses are often deemed necessary in order to avoid loosing relevant information when studying complex systems (especially when little is known about it \textit{a priori}). Nonetheless, dealing with high-dimensional datasets, obtained either by combining many different descriptors or by using intrinsically high-dimensional descriptors, is not always advantageous.\cite{Lionello2025} In many cases, the different dimensions contain projections of the same physical event, and the recombination/compression of them can lead to technical problems and information loss. For this reason, as recently demonstrated, in most cases it may be more effective to screen all dimensions/variables in the dataset searching the most informative one(s), and analyze it(them) at a deeper level, rather than combining all components using, for example, dimensionality-reduction approaches (especially considering that many components may contain negligible/redundant information alongside non-negligible noise, which can sum up affecting the analysis). In particular, this is often the case for high-dimensional dataset obtained using inherently high-dimensional descriptors (as, e.g., the Smooth Overlap of Atomic Positions (SOAP)\cite{Bartk2013soap}), where the different descriptor's components/dimensions often contain just projections of the same physical events.\cite{Lionello2025} 
In this light, the possibility to rely on a robust method that uses an absolute metric to rank data in high-dimensional datasets and identify the information-richer dimensions is particularly useful.

Here, as a representative example, we use the Smooth Overlap of Atomic Positions (SOAP)\cite{Bartk2013soap} descriptor computed on the same MD simulation of ice and water coexistence of Figure \ref{Fig3}a.
In short, the SOAP descriptor provides a high-dimensional description of the structural order/disorder, symmetry, density, etc., in the displacement of the neighbors around each molecule in the system (Figure \ref{Fig4}a, left).\cite{Bartk2013soap} In this case, the SOAP descriptor was computed using $\mathrm{n_{max}=8}$ and $\mathrm{l_{max} = 8}$, yielding 576-dimensional SOAP power spectra (Figure \ref{Fig4}a , right). 
The SOAP power spectra were calculated for each of the 2048 water molecules in the simulated system at each of the 1250 frames in the MD trajectories. Given that SOAP spectra are known to suffer from considerable local noise, in the data of Figure \ref{Fig4}, we reduced it with the spatial-averaging denoising approach\cite{Donkor2024spatavg} (obtaining $\langle SOAP \rangle$ power spectra), as done for the other descriptors in Figure \ref{Fig3} (see Methods section for complete details).\cite{Lionello2025,Martino2025,Donkor2024spatavg} The $\langle SOAP \rangle$ dataset analyzed herein thus contains $2.56 \times 10^6$ $\langle SOAP \rangle$ spectra that are 576-dimensional. 

Since in similar analyses, one typical problem is that most of these dimensions are essentially useless,\cite{Lionello2025} we treated each of these 576 $\langle SOAP \rangle$ components/dimensions as data with own distributions. We then used the MInA workflow to analyze and rank them, and to understand which ones contain more information (analogous MInA analysis for the corresponding raw -- prior to denoising -- SOAP dataset has also been conducted, and the results are provided in Supplementary Figure~\ref{FigS5}).

Recently, it was demonstrated that methods based on the variance of the data along each dimension may fall short in capturing the relevance of the information contained in the SOAP vectors (and in their components).\cite{Lionello2025} For example, \citet{Lionello2025} demonstrated that components with negligible weight in the variance (e.g., component $\langle 237 \rangle$) may be in reality highly informative (e.g., when analyzed as a time series data via single-point Onion clustering\cite{Becchi2024onion}). Nonetheless, the computational cost and complexity of analyzing all 576 SOAP components as timeseries are prohibitive, making it practically impossible to systematically screen the information contained in all components with such approaches. Being fast, scalable, and interpretable, MInA offers clear advantages to this purpose.

From the time-series of each $\langle SOAP \rangle$ component, we obtained their distributions (Figure \ref{Fig4}b) reports the KDEs of a few representative ones), which were then analyzed with MInA obtaining the data microstructure diagrams of Figure \ref{Fig4}c, and the Information-Microstructure plots of Figure \ref{Fig4}d-f.

The data microstructure diagrams of Figure \ref{Fig4}c show the populations of the different Gaussian and non-Gaussian environments that compose the data. It is worth noting that these are shown without detail of the information content of each microdomain for visual simplicity (diagrams complete with such information breakdown data are provided in Supplementary Figure~\ref{FigS6}), while these numbers are discussed here in the following.
In particular, components $\langle 197 \rangle$ and $\langle 237 \rangle$, characterized by a similar distribution, both contain in their microstructure $\Delta I=1.53~[bit]$ of information, distributed within their microdomains in a similar way: the first Gaussian microenvironments $g1$ retain $\sim 54-57\%$ of information, the second $g2$ ones $\sim31-33\%$, and the non-Gaussian $ng$,  $\sim13\%$ of information. In comparison, components $\langle 63 \rangle$ and $\langle 328 \rangle$ contain overall less information, $\Delta I = 1.21 ~[bit]$ and $\Delta I = 1.02~[bit]$ respectively, and are characterized by wider and closer Gaussian microenvironments (Figure \ref{Fig4}b). 

The information-microstructure plots of the $\langle SOAP \rangle$ components are reported in Figure \ref{Fig4}d-f. The trends are similar to those seen in Figure \ref{Fig2} and Figure \ref{Fig3}, but here the larger number of data points allows a clearer interpretation. 
The non-linear relationship between $\Sigma\mathrm{\textit{guf}}$ and $\Delta I$ (Figure \ref{Fig4}d,e) confirms that $\Sigma\mathrm{\textit{guf}}$ emphasizes more the differences among the most informative components and less those among the least informative (and least interesting) ones, whereas $\Delta I$ trends to do the opposite. 
Therefore, $\Sigma\mathrm{\textit{guf}}$ can be generally considered more effective in ranking data based on the information contained in their microstructure (although examining it alongside $\Delta I$ remains nonetheless informative). Since $\Sigma\mathrm{\textit{gef}}$ and $\Sigma\mathrm{\textit{guf}}$ are inversely related, the region where the most informative components fall differs between the two planes: they lie on the upper-left corner of the $\Delta I$-$\Sigma\mathrm{\textit{gef}}$ plane, whereas they appear in the upper-right corner of the $\Delta I$--$\Sigma\mathrm{\textit{guf}}$ plane (Figure \ref{Fig4}d-f). Read together, these two plots again demonstrate how the most informative datasets are those with substantial contributions from both Gaussian and non-Gaussian parts. Microstructurally, the different data/components can be organized according to the number of Gaussian peaks they contain (Figure \ref{Fig4}e and Figure \ref{Fig4}f, insets: orange vs. black clusters of components). A larger number of Gaussian microenvironments generally corresponds to a greater overall information content ($\Delta I$), while the precise position of a component within each plane reveals how its information is partitioned between the Gaussian and non-Gaussian microdomains. Comparing these plots is itself highly informative, since the differences in how the data are displaced between them indicate where the information is located in a given dataset: for example, whether predominantly in its Gaussian or non-Gaussian microdomains.

In Figure \ref{Fig4}e,f, the information-richest $\langle SOAP \rangle$ components in absolute terms are $\langle 197 \rangle$ (teal) and $\langle 237 \rangle$ (light green), together with the components reaching the highest values of $\Delta I$ and $\Sigma\mathrm{\textit{guf}}$, for example component $\langle 3 \rangle$ (black), component $ \langle 485 \rangle$ (orange), and component $\langle 63 \rangle$ (purple). Component $\langle 63 \rangle$ provides an instructive counter-example: its $\Sigma\mathrm{\textit{guf}}$ is $\sim20\% $ lower than that of the most informative components (e.g., $\langle 197 \rangle$ and $\langle 237 \rangle$), even though its variance is roughly a million times larger than that of component $\langle 237 \rangle$.\cite{Lionello2025} It is also worth noting how these results are consistent with recent Onion clustering-based results, which suggested $\langle 237 \rangle$ as a highly informative component in such $\langle SOAP \rangle$ dataset.\cite{Lionello2025} However, in this case, MInA unambiguously demonstrates this as it allows us to systematically screen all components and rank all of them based on their information content. 

In Figure \ref{Fig4}d,e, the most descriptive components lie in the top-right corner (high $\Delta I$ and high $\Sigma\mathrm{\textit{guf}}$) and belong to the two-peak domain (Figure \ref{Fig4}e, inset: in orange). In particular, the most informative of all ($\langle 197 \rangle$) combines two well-marked peaks and the highest measured $\Delta I$ in combination with a relatively large non-Gaussian environment and relatively low $\Sigma\mathrm{\textit{gef}}$ (Figure \ref{Fig4}e,f), meaning that the information that it contains are not only contained in its Gaussian microenvironments but also in the non-Gaussian one. Consistent results and trends are obtained also for the non-denoised SOAP dataset, where the group of most descriptive components contain also the $\langle SOAP \rangle$ highlighted in Figure \ref{Fig4}d,e (see Supplementary Figure~\ref{FigS5}). 

The results presented in Figure \ref{Fig4} demonstrate how one key advantage of MInA is that it allows screening a large number of data distributions, and rank them based on metrics that also implicitly accounts for the information contained in their (non-Gaussian) transition domains, and even without explicitly considering the time correlation between the data points (in the case of timeseries). The method becomes thus easily applicable and transferable to both static data distributions as well as timeseries data, where it captures essential features implicitly and at an affordable computational cost: once the most promising/informative features are detected, these can then be eventually studied at a deeper level with more advanced algorithms to maximize information extraction.

\subsubsection{MInA as a general approach for absolute and transferable information ranking}

\begin{figure}[htbp]
    \centering
    \includegraphics[width=1\textwidth]{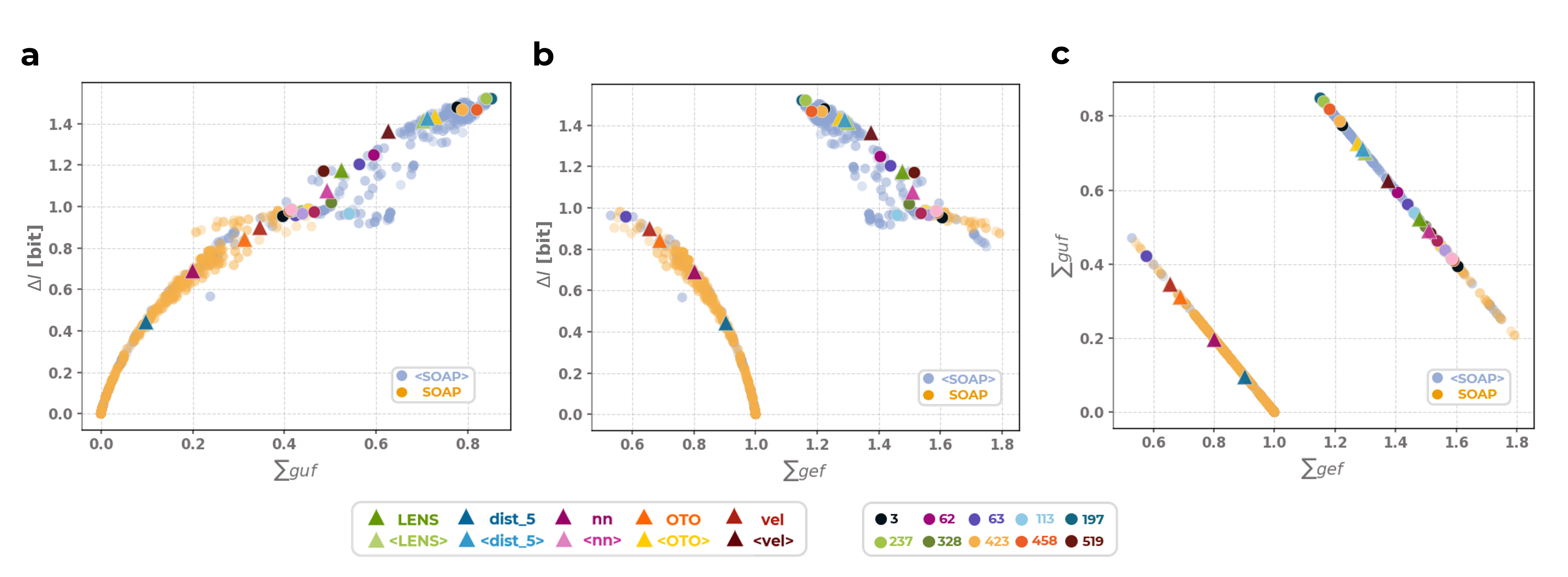}
    \caption{\textbf{General Micro-Information analysis for all the data.} (a-c) Information-microstructure plots for all dataset obtained with the different descriptors employed above. (a) $\Sigma\mathrm{\textit{guf}}$-$\Delta I$ plot, with all $D$, $\langle D \rangle$, $SOAP$, and $\langle SOAP \rangle$ dataset compared altogether. Yellow circles represent the raw SOAP components, light-clue circle the denoised $\langle\mathrm{SOAP}\rangle$ components. Colored triangle mark the heuristic descriptors (\textit{LENS}, \textit{$dist\_5$}, \textit{nn}, \textit{OTO}, \textit{vel}, together with their time-averaged counterparts $\langle\mathrm{\textit{LENS}}\rangle$, $\langle\mathrm{\textit{$dist\_5$}}\rangle$, $\langle\mathrm{\textit{nn}}\rangle$, $\langle\mathrm{\textit{OTO}}\rangle$, $\langle\mathrm{\textit{vel}}\rangle$) in lighter colors. Colored circles highlight ten representative most descriptive $\langle SOAP \rangle$ components (those highlighted in Figures \ref{Fig4}: $\langle 3 \rangle$, $\langle 62 \rangle$, $\langle 63 \rangle$, $\langle 113 \rangle$, $\langle 197 \rangle$, $\langle 237 \rangle$, $\langle 328 \rangle$, $\langle 423 \rangle$, $\langle 458 \rangle$, $\langle 519 \rangle$). (b,c) $\Sigma\mathrm{\textit{gef}}$-$\Delta I$ (b) and $\Sigma\mathrm{\textit{gef}}$-$\Sigma\mathrm{\textit{guf}}$ (c) plots with the same color and marker code.}
    \label{Fig5}
\end{figure}

Methods to compare and rank data based on their information content typically rely on measurements of their similarity. Examples have recently been reported based, e.g., on the construction of an entire high-dimensional dataset mixing all data to be compared together (e.g., via PCA, etc.) and measuring their adjacency/distance in the space of the data,\cite{Martino2025,Capelli2021,Gardin2022,Jolliffe2016_pca,Molgedey_TiCA,McInnes2018} or on information imbalance\cite{info-imbalance,Donkor2023,Wild2024}, which can reveal how much information contained in one dataset is also contained in, or lacking from, another. 
Such approaches are all based on the development of a data-driven metric, which is relative and depends on the data within the dataset\cite{Martino2025,info-imbalance} and while effective for many applications, possess intrinsic limitations. 
For example, data-driven metrics are also inherently non-transferable to data different from those used to generate them: the addition of even a single new data point that one may want to compare to the original dataset requires the entire metric to be recomputed from scratch, as the reference distances or rankings depend on the composition of the dataset as a whole. At the same time, such relative metrics also lose their resolving power as the number of compared data points grows, since pairwise differences become progressively less distinguishable.
In contrast, the comparisons made with MInA (Figures \ref{Fig2}-\ref{Fig4}) use a purely absolute and transferable metric to this end, offering a general, physically-grounded, and absolute method with some key advantages. First, it allows the comparison of a large number of data points without losing the resolution needed to capture their differences as their number increases. Second, the addition of further data to the dataset does not require recomputing the entire analysis, but only to evaluate the microstructural diagram related to the new-added data. Third, the outcome is interpretable: thanks to the microscopic decomposition of the information content, in once, MInA also reveals where most information is nested, providing precious insights on the physics of the system under study. 

To prove the transferability and potential of our method in ranking and comparing different types of high-dimensional datasets based on their information content, we compared altogether all dataset obtained using the various descriptors of Figure \ref{Fig3}, the $\langle \mathrm{SOAP} \rangle$ descriptor of Figure \ref{Fig4}, and the raw $SOAP$ descriptor of Figure~\ref{FigS5}, without recomputing the microstructural diagrams. Figure \ref{Fig5} shows the comparison in terms of information and data microstructure of all 1162 dataset (10 $D$, plus 10 $\langle \mathrm{D} \rangle$ heuristic descriptor dataset, plus 576 $\langle \mathrm{SOAP} \rangle$ and 576 $SOAP$ components). As clearly seen in Figure \ref{Fig5}, the same trends are consistently observed across variables with completely different origin, nature, and and metrics, without a need for building a global shared dataset containing all of them. 

The datasets richest in internal information are those exhibiting the largest number of clearly detectable Gaussian microenvironments, together with the largest possible $\Sigma\mathrm{\textit{guf}}$. In Figure \ref{Fig5}a,b, thus, the most informative datasets are those located in the top-right corner of the $\Delta I$-$\Sigma\mathrm{\textit{guf}}$ plane and, at the same time, in the upper-left region of  the rightmost cluster in the $\Delta I$-$\Sigma\mathrm{\textit{gef}}$ plane.
By looking more closely at the information-microstructure plots of Figure \ref{Fig5}a-b, the first visible difference is the effect of denoising on the SOAP spectra: the denoised $\langle \mathrm{SOAP} \rangle$ components (in cyan), characterized by narrower and more separated peaks, retain more information than the raw ones (in yellow), whose peaks are topically broader and sometimes missing. 
At the same time, the heuristic descriptors (depicted as triangles), which are designed \textit{ad hoc} for the study of water, and are known to perform remarkably well for aqueous systems, used here as references, fall within the cluster of the most descriptive quantities, close to $\langle \mathrm{SOAP} \rangle$ component $\langle 237 \rangle$, one of the most descriptive components, according to our metric. Physically, these are the variables able to resolve two distinct ice-water interface environments, corresponding to the ice-to-water and water-to-ice contacts as shown in previous works. \cite{Martino2025,becchi2025mine,Lionello2025}. 
Figure \ref{Fig5}c further shows that raw SOAP, denoised SOAP, and the heuristic descriptors all collapse onto the same two branches in the $\Sigma\mathrm{\textit{guf}}$-$\Sigma\mathrm{\textit{gef}}$ plane, corresponding to single-peak and two-peak distributions, and obey the same inverse relationship between the two cumulative fractions.

Taken together, these results show the efficiency of our method and highlight its advantages. The MInA approach can be easily applied, at low computational cost, to high-dimensional datasets of virtually any kind, whether synthetic, generated from MD simulations, or measured from observation of real systems, and whether derived from time-series or from static data distributions. The output of such simple and rapid analyses takes the form of plots that immediately reveal where the information resides within the data. The objective metrics used by the MInA method, allow to compare datasets of any type without embedding them into a single large dataset, and without degrading the efficiency as the number of data grows.

\section{Conclusion}

The study of complex systems and phenomena is very often dependent of the efficiency in extracting information from the data extracted from their observation. However, quantifying how much information is contained in the data, and where exactly, is most often non-trivial, especially when dealing with high-dimensional datasets composed of many different types of data that are difficult to compare to each other. 
In this work, we described a new method that allows quantifying and mapping the information contained in the microstructure of data distributions of any type using objective, transferable and quantitative metrics, which we call "\textit{Micro-Information Analysis (MInA)}". The method builds on a simple consideration: the information contained in the data is tightly related to features of their distribution -- such as, e.g., the number of density peaks, their sharpness/broadness, separation, and the non-Gaussian residual part of the data -- which are evident at a glance (and even before using advanced analysis approaches), and that we refer to as the dataset microstructure. 

The central premise of MInA is that both Gaussian and non-Gaussian microenvironments that can be identified in the data contain certain information of their own, which, in the case of data referring to physical systems, also have different physical interpretability. 
While distinct Gaussian microdomains correspond to different microstates in the system the data refer to, the non-Gaussian part of a dataset is not merely noise, but it may conceal physically relevant information on transition patterns, dynamical interconversions between microstates, etc.\cite{Becchi2024onion,becchi2025mine} 
Accordingly, the presence in the data of a large number of distinct Gaussian and non-Gaussian microdomains indicates that more information is nested in them than that captured by their global Shannon entropy (often ambiguous): namely, the information contained in their microstructure. 

Leveraging such concepts, the MInA method performs a preliminary data clustering discovering, in exquisitely microscopic fashion, all distinct Gaussian microdomains -- hierarchically proceeding from those identified by the highest to the lowest density peaks  --, until no further peaks can be found, and assigning the remaining data to a residual non-Gaussian microcluster. MInA thus implicitly performs a physically-grounded (rather than, e.g., merely numerical) clustering of the data, subdividing the data in terms of clearly distinguishable, physically-different microstates. 
At the same time, by exploiting the clustering-and-minimization of Shannon entropy principle\cite{becchi2025mine} -- which relies on the reduction of Shannon entropy following clustering of data into distinct microclusters --, it quantifies the microscopic-level information contained in each Gaussian and non-Gaussian microcluster. In doing this, MInA effectively quantifies the information gain associated with unveiling (and thus contained into) the data microstructure. This information is quantified in $[bit]$ and is directly related to the data microstructure. Such information-microstructure relationships are directly mapped and represented by MInA in the form of data microstructure diagrams and information-microstructure plots that are quantitative and easy to interpret.

Beyond ranking large collections of data from the most to the least informative, MInA also indicates where the additional information of one dataset with respect to another resides, i.e., whether in its Gaussian or in its non-Gaussian microdomains.
Because the resulting metric is physically grounded and absolute rather than relative, the information content of dataset of arbitrarily different nature can be compared without the need to embed them into a common high-dimensional space in which their relative distances are measured, an operation that would otherwise render the ranking of large amounts of data progressively less sensitive and efficient. Another advantage of MInA lies in the physical interpretability of the results: the representation of the microstructure of the data, and its relation to the information contained within it, is readily interpretable.

Here we have tested the effectiveness of MInA across different types of datasets. The method shows the ability to efficiently and easily handle dataset of millions of points and more than one thousand dimensions on a standard computer, allowing to rank data having completely different origin, nature, and metrics, based on a single, robust, transferable, and physically-grounded (information-entropy based) scale.
In general, all results reported herein demonstrate how a growing number of Gaussian microenvironments is typically associated to a higher information content of the data and that, for a fixed Gaussian content (data containing the same number of Gaussian microclusters), a larger non-Gaussian fraction corresponds to a more informative dataset.
The analysis performed by MInA is completely unsupervised and yields several parameters that map directly onto well-defined physical features. 
In particular, the cumulative Gaussian unexplained fraction  $\Sigma\mathrm{\textit{guf}}$ -- which encodes information not only on how many Gaussian microclusters are present in the data, but also how much data remain out of them, being part of the non-Gaussian domain -- exhibits a clear relationship with the information content of the data $\Delta I$ (the information contained in the data microstructure). This emerges as a single scalar descriptor that is more effective than global quantities such as, e.g., Shannon entropy or negentropy, for ranking datasets according to their informativeness, even when the data are completely different from each other (see, e.g., Figure \ref{Fig5}). 

MInA provides a completely unsupervised and yet robust and physically-interpretable data-driven approach useful for the study of complex systems from their data, where the identification of where information is located in the data microstructure may unveil previous insights into the physical microstates that dominate the system under study. Given its flexibility and simplicity, we believe that the MInA method will be particularly useful for high-dimensional data analysis in a broad sense, as well as for data science in general.

\newpage

\section{Methods}
\subsection{Water/ice simulation}

The trajectory of the atomistic ice-water system investigated in this study are taken from previous work.\cite{Caruso2023,Crippa2023,Becchi2024onion,becchi2025mine,Lionello2025,Martino2025,Doria2025} The simulated ice/water system consists of 2048 water molecules modeled with the TIP4P/Ice potential.~\cite{Abascal2005} The initial configuration of the simulation box consists of with 50\% (1024) of the molecules arranged in the hexagonal ice Ih phase and 50\% (1024) in the liquid phase.\cite{Matsumoto2017} The phase coexistence molecular dynamics (MD) simulation was performed at the melting temperature (268~K for the TIP4P/Ice model).\cite{Abascal2005} Temperature was controlled using the v-rescale thermostat with a relaxation time of 0.2~ps. The system was equilibrated for 10~ns under ambient pressure using the c-rescale barostat\cite{Bernetti2020_barostat}, with a water compressibility of $4.5 \times 10^{-5}$~bar$^{-1}$. Both equilibration and production runs were carried out in the NPT ensemble under semi-isotropic conditions using the GROMACS simulation package.\cite{Abraham2015} Pressure coupling was applied only along the axis perpendicular to the ice/water interface. The production run lasted 50~ns, with configurations saved every 1~ps. The trajectory analyzed with a sampling interval of 40 ps. Additional details on the MD protocol can be found in Ref.~\citenum{Caruso2023}

\subsection{Computation of the descriptors}

The Local Environment and Neighbor Shuffling (\textit{LENS}) descriptor\cite{Crippa2023} is a dynamic descriptor that quantifies how rapidly the local neighborhood of a particle is reshuffled over time. For each particle, it compares the list of neighbor identities found within a cutoff radius $r_{\mathrm{cut}} = 10$~\AA. \cite{Doria2025,Martino2025} \textit{LENS} was computed using the \texttt{Dynsight} package.\cite{Martino2026}

The distance to the fifth neighbor \textit{(dist\_5}) is a static descriptor commonly used to characterize local environments in water. For each particle, is the instance from its center to the fifth closest particle. 

The number of neighbors (\textit{nn}) is the number of particles lying within a cutoff radius of the central particle, here set to $r_{\mathrm{cut}} = 10$~\AA. 

The Orientational Tetrahedral Order (\textit{OTO}) parameter measures how closely the local arrangement of a particle and its neighbors resembles a perfect tetrahedron. It is computed from the angles subtended at the central particle by pairs of these neighbors, and equals 1 for an ideal tetrahedral environment, decreasing towards 0 as the geometry departs from tetrahedrality. The velocity (vel) is a static, single-frame descriptor given by the modulus of the instantaneous velocity of each  particle.

The Smooth Overlap of Atomic Positions (\textit{SOAP}) descriptor\cite{Bartk2013soap} was computed for each oxygen atom in the simulated system, centered on its position and including all neighboring atoms within a cutoff radius $r_{\mathrm{cut}} = 10$~\AA. At each sampled simulation step, the atomic neighborhood was thus represented as a Gaussian-smoothed local density function, and the SOAP power spectrum was obtained by projecting this density onto the chosen set of radial and angular basis functions characterized by the parameters $n_{\text{max}} = 8$ and $l_{\text{max}} = 8$. These parameters were selected to ensure an accurate and converged description of the structural environment while maintaining computational efficiency. The resulting 576-dimensional \textit{SOAP} vectors were used as local structural fingerprints for subsequent analyses, including non-Gaussianity assessment and information-based clustering. \textit{SOAP} power spectra were computed using the DScribe~\cite{Himanen2020,Laakso2023} and \texttt{Dynsight} packages\cite{Martino2026}.

Denoised versions of each descriptor were obtained via spatial averaging\cite{Donkor2024spatavg,Martino2025,Lionello2025}: the value of a descriptor for each particle was replaced by the average over the particle itself and all its neighbors within a cutoff radius  $r_{\mathrm{cut}} = 10$~\AA. The calculation was performed using the \texttt{Dynsight}\cite{Martino2026} package.

\subsection{Computation of Micro-Information Analysis}
The distribution of each dataset analyzed herein was estimated with a Gaussian kernel density estimator (KDE), implemented in \texttt{seaborn.kdeplot}~\cite{Waskom2021_seaborn}. The KDE was evaluated on the default regular grid of 200 points, spanning the data range extended by three bandwidths on each side, and its bandwidth was scaled by a factor \texttt{bw\_adjust} chosen for each class of datasets (see below). The total area under the KDE, $A_{\mathrm{KDE}}$, was obtained by trapezoidal integration,
\begin{equation}
    A_{\mathrm{KDE}} = \int \mathrm{KDE}(x)\,\mathrm{d}x .
    \label{eq:kde-area}
\end{equation}
Local maxima of the KDE were identified with the \texttt{scipy.signal.find\_peaks} routine~\cite{2020SciPy-NMeth}, using a small prominence threshold to retain weak secondary features. Only maxima whose height exceeds a fraction $\tau = 0.05$ of the global maximum were considered significant. The $n$ significant maxima were then ranked in descending order of KDE height, and one Gaussian was fitted to each of them, following this order.

For each maximum $k$, located at $x_k^{*}$, a local fitting window was defined by scanning outward from $x_k^{*}$ on both sides for as long as the KDE does not increase, i.e., up to the nearest local minimum on each side. An initial estimate of the width was obtained from the full width at half maximum (FWHM) of the peak,
\begin{equation}
    \sigma_{k}^{0} = \frac{x_{\mathrm{right}} - x_{\mathrm{left}}}{2\sqrt{2\ln 2}} ,
    \label{eq:sigma0-fwhm}
\end{equation}
where $x_{\mathrm{left}}$ and $x_{\mathrm{right}}$ are the abscissae at which the KDE first falls to half of its value at $x_k^{*}$, moving outward from the maximum, so that $x_{\mathrm{right}} - x_{\mathrm{left}}$ is the FWHM. A Gaussian of the form
\begin{equation}
    G_k(x) = h_k \exp\!\left(-\frac{(x-\mu_k)^2}{2\sigma_k^2}\right)
    \label{eq:gaussian-fit}
\end{equation}
was then fitted to the KDE values within the window by nonlinear least squares
\\
(\texttt{scipy.optimize.curve\_fit})~\cite{2020SciPy-NMeth}, with $h_k$, $\mu_k$ and $\sigma_k$ as free parameters, initialized to 
\\
$\mathrm{KDE}(x_k^{*})$, $x_k^{*}$ and $\sigma_k^{0}$, respectively. The fit was constrained to $h_k \geq 0$, to $\mu_k$ within the KDE grid, and to $\sigma_k$ not exceeding the grid range; if the optimization did not converge, the initial estimates were retained.

The deviation from Gaussianity was quantified through an iterative union-overlap procedure, in which one Gaussian is added at each step. After the $k$-th Gaussian is fitted, its pointwise overlap with the KDE is accumulated into a running union-overlap profile.
The area under $u_k$ is the portion of the KDE explained by the first $k$ Gaussians, and defines the Gaussian explained fraction after $k$ steps,
\begin{equation}
    \mathrm{\textit{gef}}_k = \int u_k(x)\,\mathrm{d}x ,
    \label{eq:gef-k}
\end{equation}
while the remaining area defines the Gaussian unexplained fraction,
\begin{equation}
    \mathrm{\textit{guf}}_k = A_{\mathrm{KDE}} - \mathrm{\textit{gef}}_k \simeq 1 - \mathrm{\textit{gef}}_k ,
    \label{eq:guf-k}
\end{equation}
with $\mathrm{\textit{gef}}_0 = 0$ and $\mathrm{\textit{guf}}_0 = A_{\mathrm{KDE}} \simeq 1$ before any Gaussian is fitted. Because the union-overlap profile can only grow, $\mathrm{\textit{gef}}_k$ is non-decreasing in $k$, while the unexplained fractions form a non-increasing sequence, $\mathrm{\textit{guf}}_1 \geq \mathrm{\textit{guf}}_2 \geq \cdots \geq \mathrm{\textit{guf}}_n$: each additional Gaussian accounts for a further portion of the KDE not captured by the previous ones. The first term, $\mathrm{\textit{guf}}_1$, measures the fraction left unexplained by the dominant peak alone, while the subsequent terms quantify the incremental contribution of the further Gaussian microdomains.

Two scalar summary metrics were derived from these sequences, which are directly represented in the microstructural diagrams shown in the main text. The cumulative Gaussian explained fraction is the sum of the explained fractions over all the fitting steps,
\begin{equation}
    \Sigma\mathrm{\textit{gef}} = \sum_{k=1}^{n} \mathrm{\textit{gef}}_k ,
    \label{eq:sum-gef}
\end{equation}
and the cumulative Gaussian unexplained fraction is the corresponding sum of the unexplained fractions,
\begin{equation}
    \Sigma\mathrm{\textit{guf}} = \sum_{k=1}^{n} \mathrm{\textit{guf}}_k .
    \label{eq:sum-guf}
\end{equation}
In the microstructural diagrams, where each step is represented by a bar of unit width, $\Sigma\mathrm{\textit{gef}}$ and $\Sigma\mathrm{\textit{guf}}$ correspond to the total areas of the explained and unexplained portions of the bars, respectively, and $\Sigma\mathrm{\textit{gef}} + \Sigma\mathrm{\textit{guf}} = n\,A_{\mathrm{KDE}} \simeq n$.

Finally, the Gaussian and non-Gaussian fractions of the distribution were defined as the explained and unexplained fractions at the last step,
\begin{equation}
    g = \mathrm{\textit{gef}}_n , \qquad ng = \mathrm{\textit{guf}}_n .
    \label{eq:g-ng}
\end{equation}
These quantities measure the portion of the KDE that is captured, and that cannot be accounted for, respectively, by the $n$ fitted Gaussians. For each distribution, the analysis therefore yields the number of detected peaks $n$, the cumulative metrics $\Sigma\mathrm{\textit{gef}}$ and $\Sigma\mathrm{\textit{guf}}$, and the Gaussian ($g$) and non-Gaussian ($ng$) fractions.

For the synthetic distributions, \texttt{bw\_adjust} $= 0.3$ and a prominence threshold of $3\times10^{-3}$ were used; for the single-particle descriptors and the SOAP components, \texttt{bw\_adjust} $= 2$ and a prominence threshold of $10^{-4}$.

The information contained in the microstructure was quantified via the Maximum Information Extraction (MInE) approach~\cite{becchi2025mine}, which is based on the decrease of the Shannon entropy when the data are partitioned into statistically relevant, distinct microdomains. In the present analysis, these microdomains are the Gaussian and non-Gaussian environments identified by the fit. The data range is divided by the intersections between consecutive fitted Gaussians, so that each data point has a single candidate Gaussian; the point is assigned to it if the Gaussian accounts for at least a fixed fraction of the local KDE value, and to a residual, non-Gaussian cluster otherwise.

Denoting by $w_c$ and $H_c$ the population fraction and the Shannon entropy of each cluster $c$ (the $n$ Gaussian clusters and the residual one), the entropy of the partitioned data is
\begin{equation}
    H_{\mathrm{clust}} = \sum_c w_c H_c ,
    \label{eq:h-clust}
\end{equation}
where all entropies are computed in bits from histograms with $N_{\mathrm{bins}} = 200$ uniform bins spanning the whole data range. The results are robust with respect to this choice, as shown in Figure~\ref{FigS7}, where the analysis is repeated with 50, 1000 and 10000 bins.  
Owing to the concavity of the Shannon entropy, $H_{\mathrm{clust}}$ is always lower than or equal to the entropy $H_0$ of the unpartitioned data, with $H_{\mathrm{clust}} = H_0$ when no microdomains are detectable. The difference between the two equals the information obtained by uncovering the microstructure of the data,
\begin{equation}
    \Delta I = H_0 - H_{\mathrm{clust}} = \sum_c \Delta I_c , \qquad \Delta I_c = w_c\,(H_0 - H_c) ,
    \label{eq:info-gain}
\end{equation}
where the second equality follows from $\sum_c w_c = 1$ and decomposes the information gain into the contributions of the individual microdomains, so that the information content of each of them can be quantified separately. The resulting $\Delta I$ depends only weakly on the number of bins: $\Delta I$ fluctuates for coarse discretizations, below \~50 bins, and converges to a stable plateau above \~100–200 bins (Figure~\ref{FigS8}). 

The same assignment also provides a population-based definition of the cumulative fractions. Adding the Gaussians one at a time in descending order of peak height, and denoting by $W_k$ the fraction of data points assigned to one of the first $k$ Gaussians, the cumulative Gaussian explained and unexplained fractions are
\begin{equation}
    \Sigma\mathrm{\textit{gef}} = \sum_{k=1}^{n} W_k , \qquad
    \Sigma\mathrm{\textit{guf}} = \sum_{k=1}^{n} \left(1 - W_k\right) .
    \label{eq:sum-gef-guf-pop}
\end{equation}
These quantities count populations of data points rather than areas under the KDE, and can therefore be directly compared with $\Delta I$, which is computed on the partition obtained with all $n$ Gaussians. This approach is general and applies equally to static distributions and time series, providing an objective, transferable metric for datasets of different nature.

%%%%%%%%%%%%%%%%%%%%%%%%%%%%%%%%

\subsection{Data availability}
Complete code, data and materials pertaining to the data and trajectories employed, and the analysis conducted herein, are available at: 
\\
\url{https://github.com/GMPavanLab/Micro-Information_Analysis} (the final version of complete data and materials pertaining to all analyses conducted herein will be available as a definitive Zenodo link upon acceptance of the final version of this paper).
The MInA plugin will be patched and made available on the DYNSIGHT\cite{Martino2026} analysis platform as soon as possible (and within acceptance of the final version of this work).  
Other information needed is available from the corresponding author upon reasonable request.

\subsection{Acknowledgments}
G.M.P. acknowledges the support received by the European Research Council (ERC) under the European Union’s Horizon 2020 research and innovation program (Grant Agreement no. 818776 - DYNAPOL).

%\subsection{Supporting Information Available}
%Supporting Information contains data related to:

\bibliography{bibliography}

%%%%%%%%%%%%%%%%%%%%%%%%%%%%%%%%%%%%%%%%%%%%%%%%%%%%%%%%%%%%%%%%%%%%%
%% SUPPORTING INFORMATION
%%%%%%%%%%%%%%%%%%%%%%%%%%%%%%%%%%%%%%%%%%%%%%%%%%%%%%%%%%%%%%%%%%%%%
\clearpage
\appendix
\renewcommand{\thefigure}{S\arabic{figure}}
\renewcommand{\thetable}{S\arabic{table}}
\renewcommand{\theequation}{S\arabic{equation}}
\setcounter{figure}{0}
\setcounter{table}{0}
\setcounter{equation}{0}

\begin{center}
{\large\textbf{Supporting Information}}\\[1ex]
{\large\textbf{Learning the information contained in the microstructure of data distributions}}
\end{center}

\begin{figure}[htbp]
    \centering
    \includegraphics[width=1\textwidth]{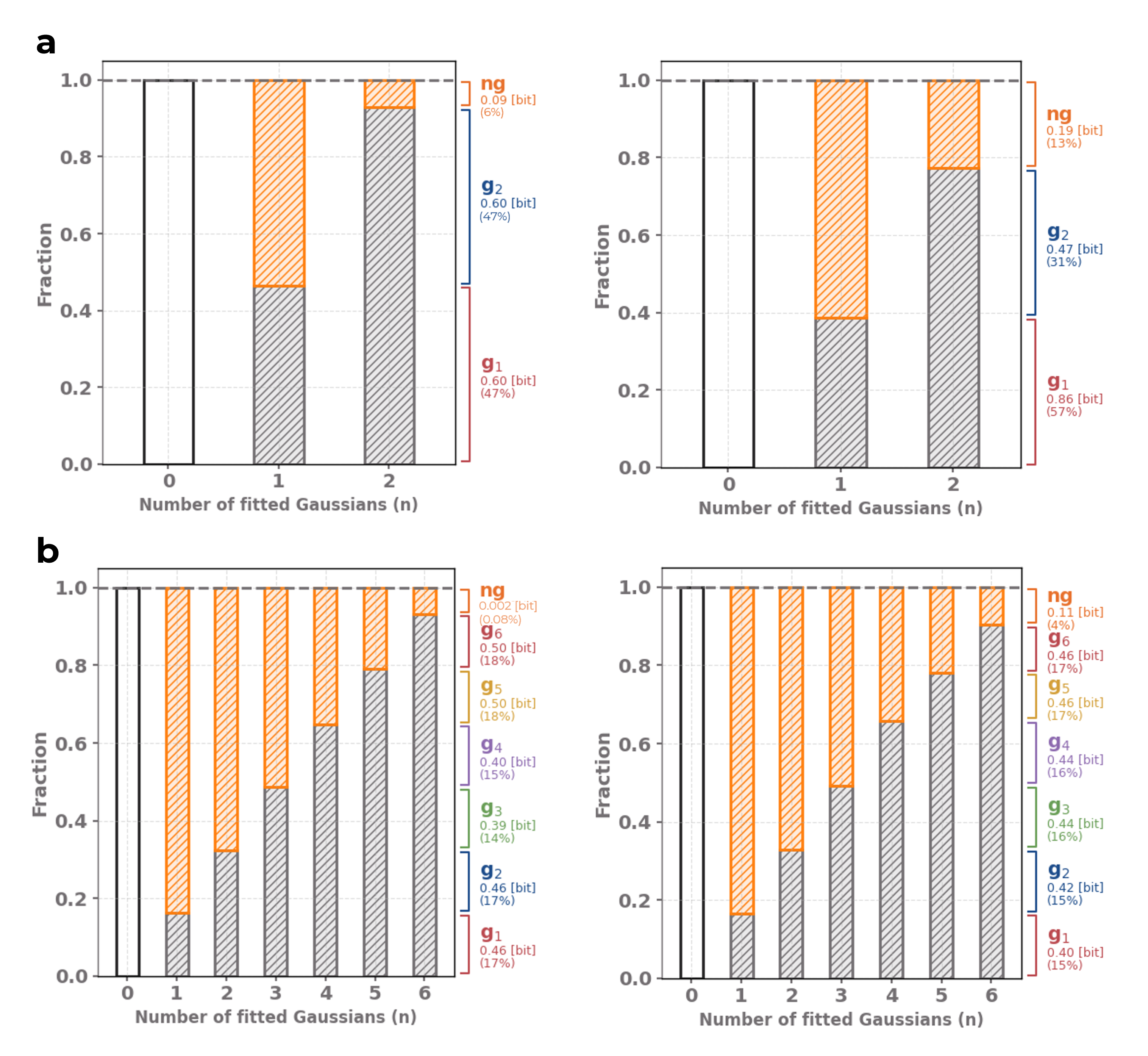}
    \caption{\textbf{Entropy breakdown detail for microstructure diagrams of Figure 1(c-f).} Data microstructure diagrams of the synthetic distributions shown in Figure 1(c-f), reported together with the information content of each Gaussian and non-Gaussian microdomain.
    (a) Distributions with two Gaussian peaks ($n=2$), corresponding to Figure 1(c) (left) and Figure 1(d) (right). The more prominent non-Gaussian region of the second distribution raises the information carried by \textit{ng} from 0.09 \textit{bit} (6\%) to 0.19 \textit{bit} (13\%) and redistributes the information among the Gaussian microdomains, which become unequal ($g_1$: 0.86 \textit{bit}; $g_2$: 0.47 \textit{bit}). (b) Distributions with six Gaussian peaks ($n=6$), corresponding to Figure 1(e) (left) and Figure 1(f) (right). In the first case the non-Gaussian microdomain carries a negligible amount of information (0.002 \textit{bit}, 0.08\%), and almost all of the information is distributed among the six Gaussian microdomains. In the second case the more pronounced non-Gaussian regions between the peaks raise the \textit{ng} contribution to 0.11 \textit{bit} (4\%).}
    \label{FigS1}
\end{figure}

\begin{figure}[htbp]
    \centering
    \includegraphics[width=1\textwidth]{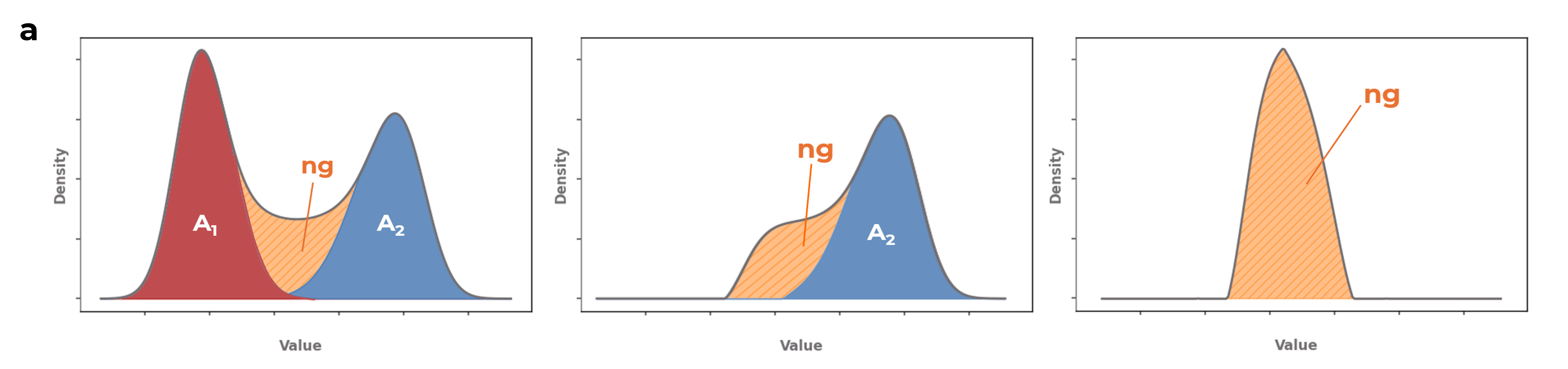}
    \caption{\textbf{Sequential deletion of the Gaussian microclusters from the example distribution of Figure 1a.} a) Detection-classification-deletion workflow, analogous to that used in Onion clustering, applied to the example distribution. Left: original distribution, decomposed into the two Gaussian microclusters $g_1$ (red) and $g_2$ (blue) and the non-Gaussian microcluster $ng$ (orange, hatched). Center: distribution remaining after the detection and deletion of $g_1$. Right: distribution remaining after the subsequent deletion of $g_2$. Once the dominant Gaussian peaks are removed, the residual $ng$ microenvironment shows a clear, sharp structure of its own, consistent with its low Shannon entropy ($H_{ng} = 2.78~[bit]$).}
    \label{FigS2}
\end{figure}

\begin{figure}[htbp]
    \centering
    \includegraphics[width=1\textwidth]{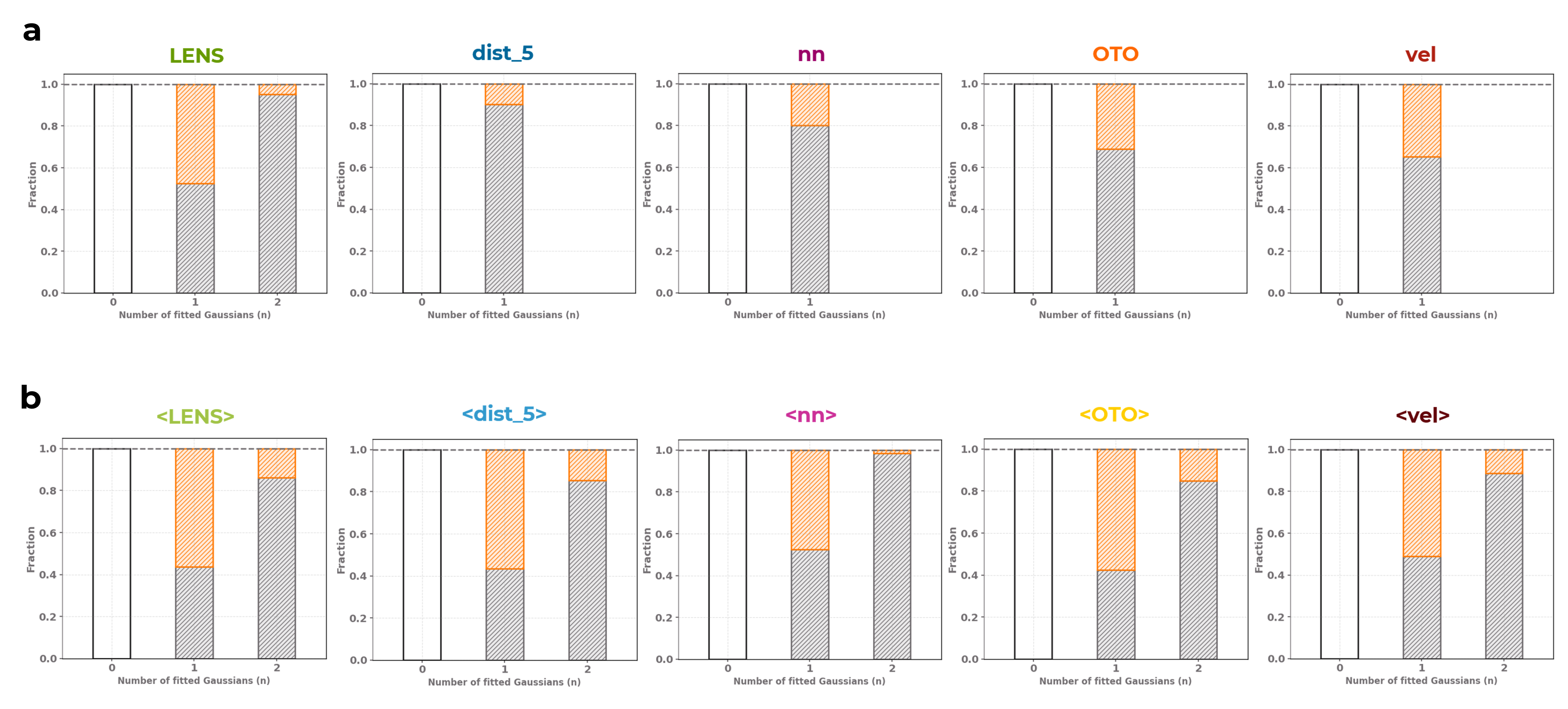}
    \caption{\textbf{Heuristic descriptors microstructure diagrams} Microstructure diagrams showing the cumulative Gaussian explained fraction $\Sigma\mathrm{\textit{gef}}$ (gray hatched region) and the cumulative Gaussian unexplained fraction $\Sigma\mathrm{\textit{guf}}$ (orange hatched region) as a function of the number of fitted Gaussian \textit{n}, for each descriptors (a) Raw descriptors (b) Corresponding denoised descriptors. The number of fitted Gaussian \textit{n} reflects the number of statistically distinct Gaussian microdomains identified for each descriptors.}
    \label{FigS3}
\end{figure}

\begin{figure}[htbp]
    \centering
    \includegraphics[width=1\textwidth]{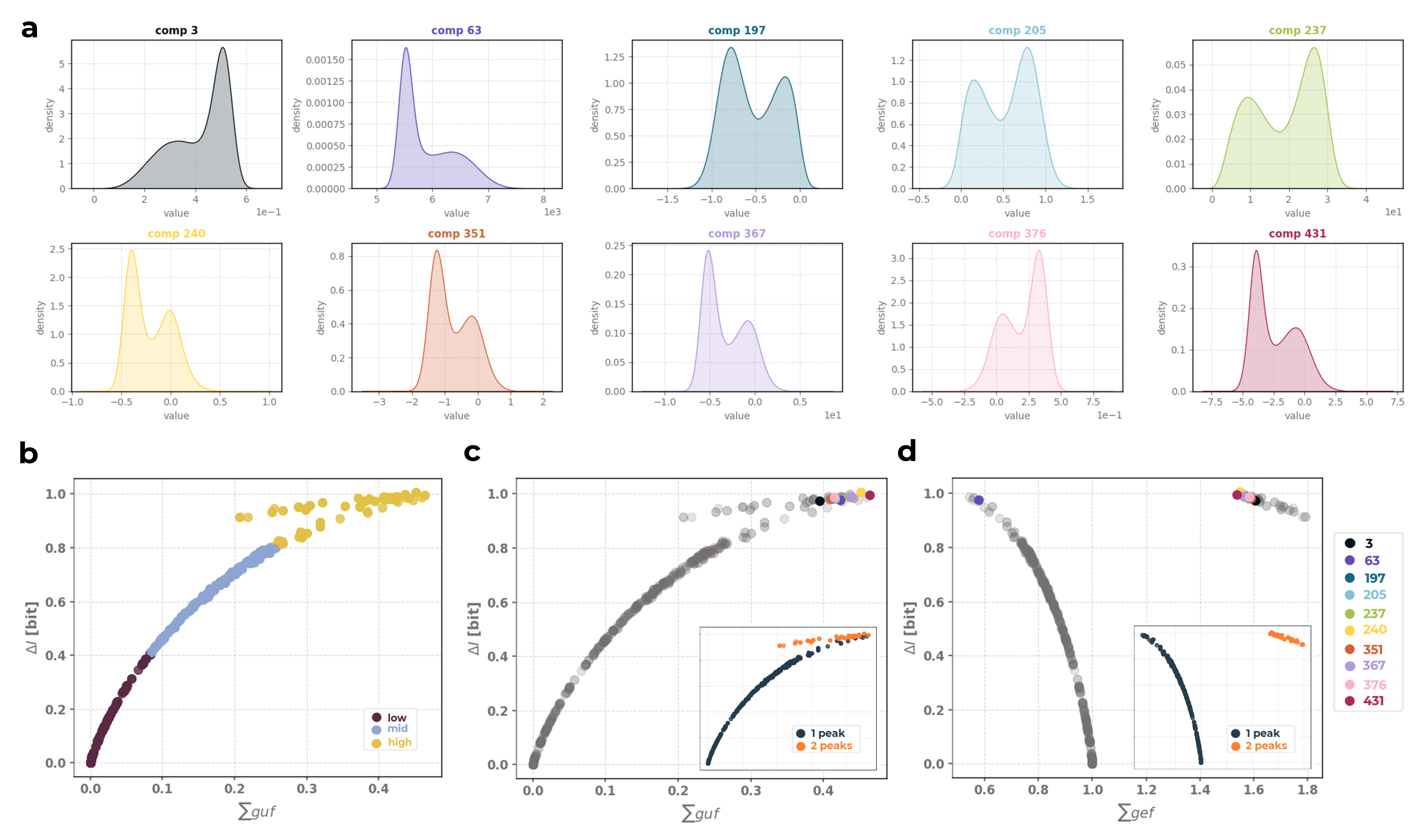}
    \caption{\textbf{Micro-Information analysis for the raw SOAP components} (a) Kernel density estimates (KDE) of ten representative raw SOAP components, illustrating the diversity of distribution shapes observed across the dataset, from markedly asymmetric, single dominant peak distribution to well separated bimodal distribution. (b-d) Structure-Information plots for the full set of raw SOAP components, with $\Delta I$ plotted as a function of $\Sigma\mathrm{\textit{guf}}$ (b,c) and $\Sigma\mathrm{\textit{gef}}$ (d). (b) Components are colored according to their informativiness into three groups, low (red), mid (blue), and high (yellow), highlighting how $\Delta I$ increase together with $\Sigma\mathrm{\textit{guf}}$. In (c) and (d), the ten representative components are highlighted in colors ; the insets sho the same data colored according to the detected number of Gaussian microdomains (1 peak, dark blue; 2 peaks, orange). Figure~\ref{FigS4} reports the microstructure diagrams of the components highlighed in (c) and (d).}
    \label{FigS5}
\end{figure}

\begin{figure}[htbp]
    \centering
    \includegraphics[width=1\textwidth]{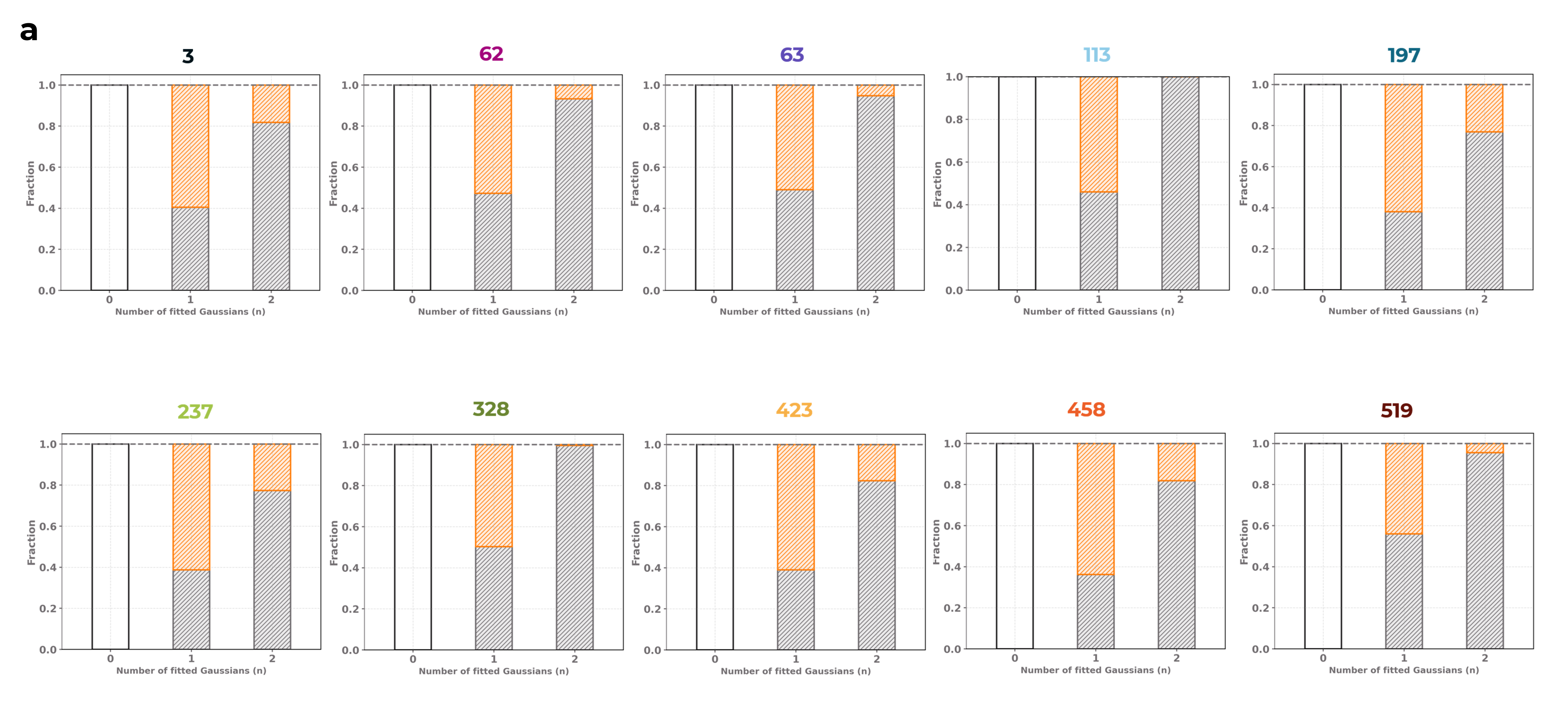}
    \caption{\textbf{Microstructure diagrams for ten representative denoised SOAP components} Micro-structure diagrams showing the cumulative Gaussian explained fraction $\Sigma\mathrm{\textit{gef}}$ (grey hatched region) and the cumulative Gaussian unexplained fraction $\Sigma\mathrm{\textit{guf}}$ (orange hatched region) as a function of the number of fitted Gaussian \textit{n}, for ten representative denoised SOAP components selected from the SOAP dataset: component 3,62,63,113,197 (top raw), and components 237,328,423,458,519 (bottom row).}
    \label{FigS4}
\end{figure}

\begin{figure}[htbp]
    \centering
    \includegraphics[width=1\textwidth]{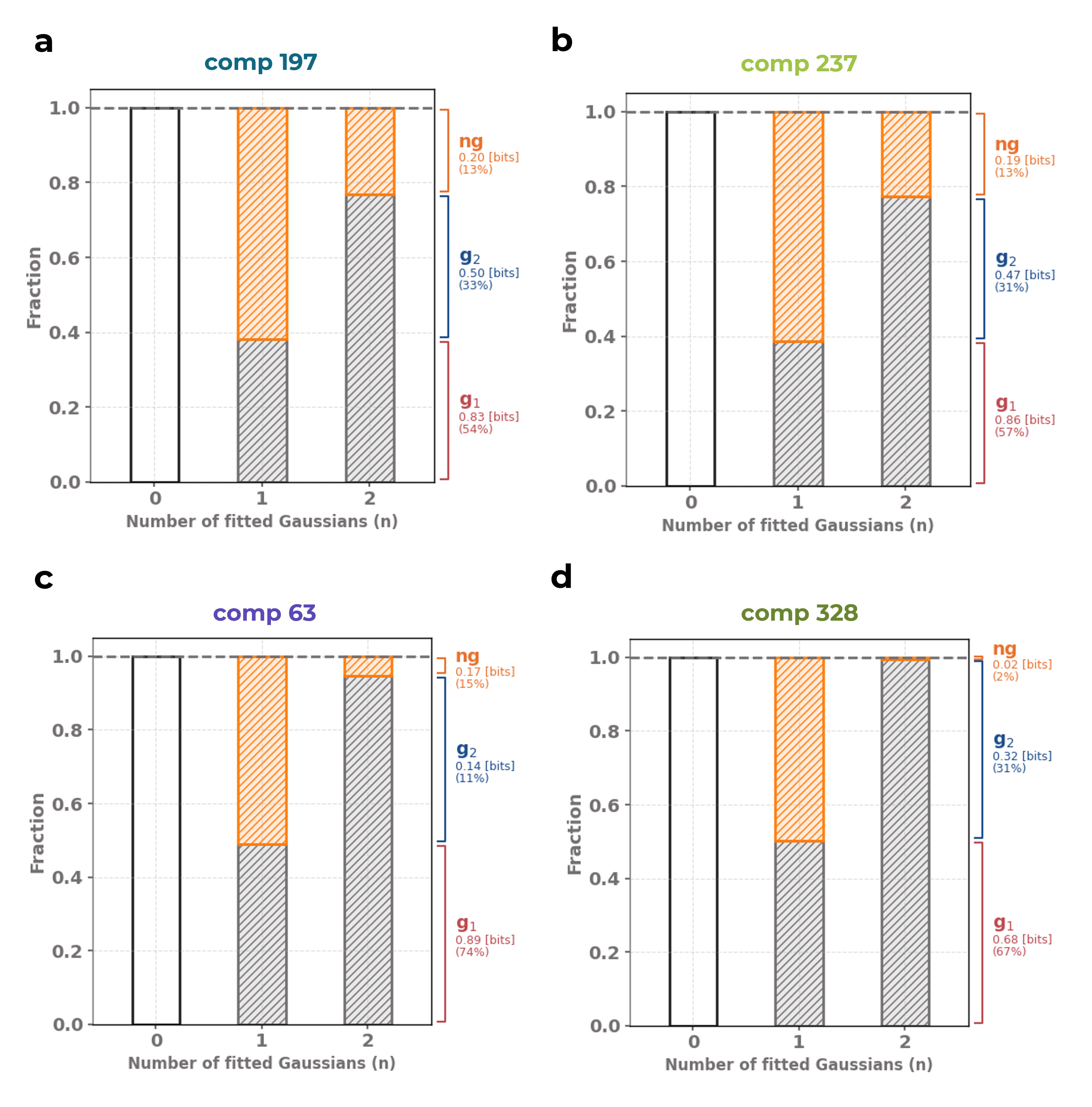}
    \caption{\textbf{Entropy breakdown detail for the microstructure diagrams of components $\langle 197 \rangle$, $\langle 237 \rangle$, $\langle 63 \rangle$ and $\langle 328 \rangle$.} Data microstructure diagrams of the four components shown in Figure Fig4, reported together with the information content of each Gaussian and non-Gaussian microdomain. (a) Component $\langle 197 \rangle$ and (b) component $\langle 237 \rangle$ have similar distributions and the same total information, $\Delta I \simeq 1.53$ \textit{bit}. This information is split in a similar way among the microdomains: $g_1$ carries 0.83 \textit{bit} (54\%) and 0.86 \textit{bit} (57\%), $g_2$ carries 0.50 \textit{bit} (33\%) and 0.47 \textit{bit} (31\%), and \textit{ng} carries 0.20 \textit{bit} and 0.19 \textit{bit} (13\% in both cases). (c) Component $\langle 63 \rangle$ and (d) component $\langle 328 \rangle$ contain less information overall, $\Delta I = 1.21$ \textit{bit} and $\Delta I = 1.02$ \textit{bit} respectively.}
    \label{FigS6}
\end{figure}

\begin{figure}[htbp]
    \centering
    \includegraphics[width=1\textwidth]{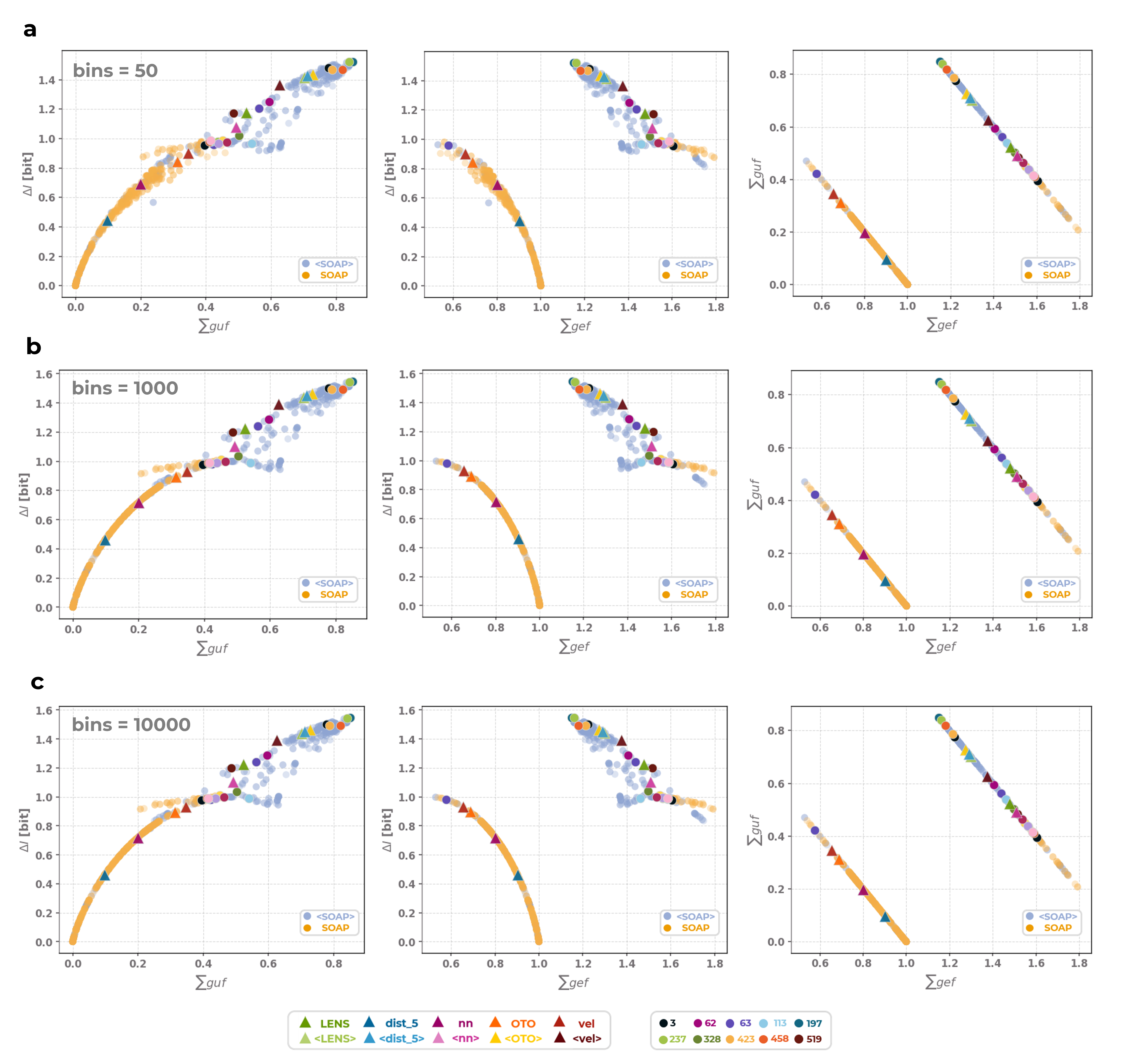}
    \caption{\textbf{Robustness of the information trends with respect to the histogram binning.} Information content $\Delta I$ and Gaussian contributions $\Sigma\mathrm{\textit{guf}}$ and $\Sigma\mathrm{\textit{gef}}$ for all the descriptors analysed, recomputed with three different numbers of histogram bins: (a) 50 bins, (b) 1000 bins and (c) 10000 bins. In each row the left panel reports $\Delta I$ as a function of $\Sigma\mathrm{\textit{guf}}$, the central panel reports $\Delta I$ as a function of $\Sigma\mathrm{\textit{gef}}$, and the right panel reports $\Sigma\mathrm{\textit{guf}}$ as a function of $\Sigma\mathrm{\textit{guf}}$, as in Figure 5 of the main text. Orange circles correspond to the raw SOAP components and light blue circles to the time-averaged $\langle$SOAP$\rangle$ components. The selected components discussed in the main text are highlighted with the colours indicated in the legend. Triangles correspond to the other descriptors (Figure 3) and their time-averaged counterparts.}
    \label{FigS7}
\end{figure}

\begin{figure}[htbp]
    \centering
    \includegraphics[width=1\textwidth]{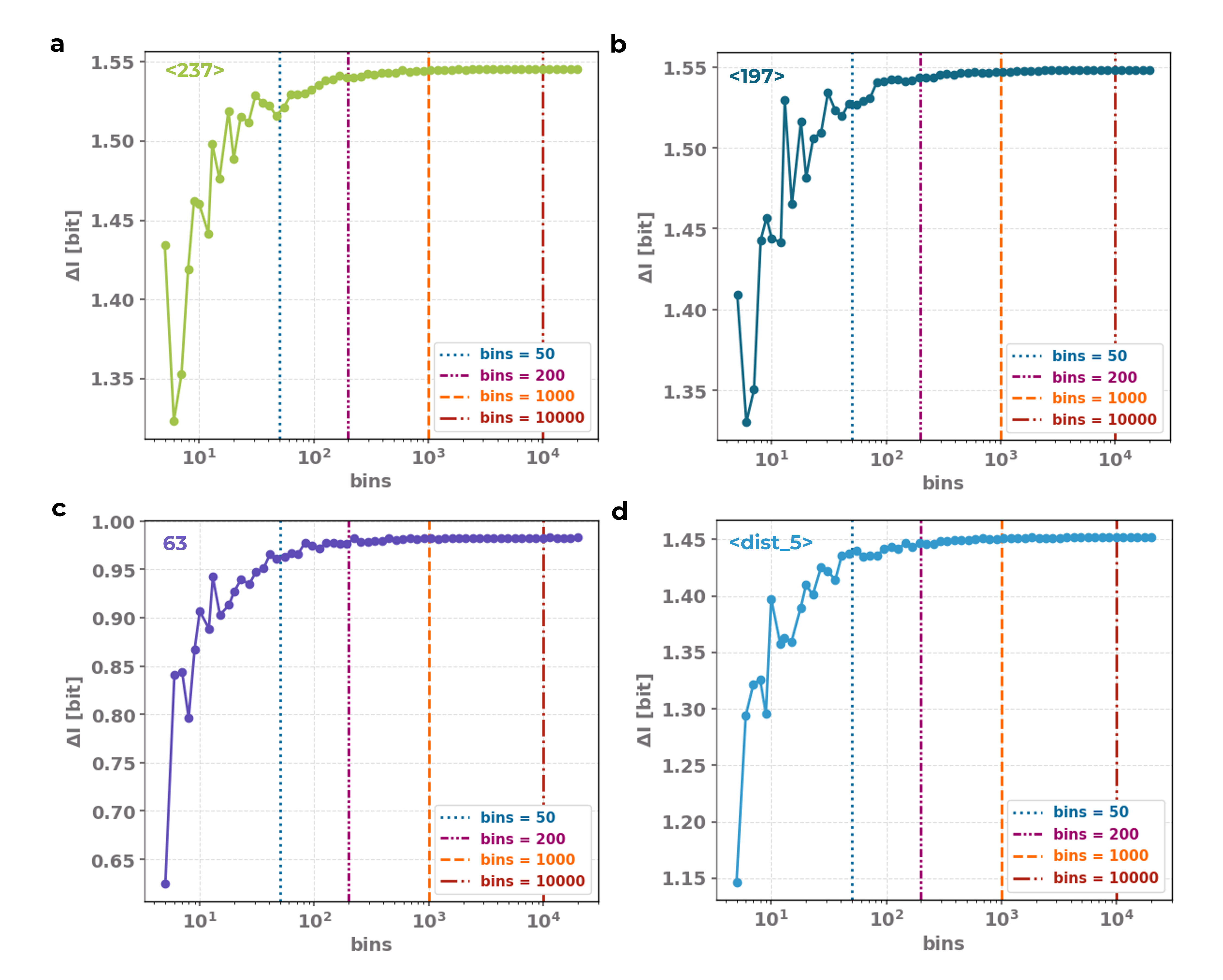}
    \caption{\textbf{Dependence of the information gain on the histogram binning.} Information gain $\Delta I$ as a function of the number of uniform histogram bins used to compute the Shannon entropies, for four representative datasets: (a) SOAP component $\langle 237 \rangle$, (b) SOAP component $\langle 197 \rangle$, (c) raw SOAP component $63$ and (d) the time-averaged descriptor $\langle$dist\_5$\rangle$. The number of bins ranges from 5 to $2\times10^{4}$ and is shown on a logarithmic scale. Vertical lines mark 50 (dotted), 200 (dash-dotted), 1000 (dashed) and 10000 (dash-dot-dotted) bins. From 200 bins (number of bins adopted in the main text) onwards it changes by less than $\sim$0.01 [bit] in all cases. In this regime the information content of the microstructure is therefore independent of the binning.}
    \label{FigS8}
\end{figure}

\end{spacing}
\end{document}